\documentclass{aa}         

\usepackage{graphicx}
\usepackage{natbib} 
\bibpunct{(}{)}{;}{a}{}{,} 

\usepackage{verbatim}
\usepackage{txfonts}
\usepackage[figuresright]{rotating}
\usepackage{url}

\begin{document}
   \title{Cumulative physical uncertainty in modern stellar models}

   \subtitle{I. The case of low-mass stars.}

   \author{G. Valle \inst{1}, M. Dell'Omodarme \inst{1}, P.G. Prada Moroni \inst{1,2}, S. Degl'Innocenti \inst{1,2}
          }

   \authorrunning{Valle, G. et al.}

   \institute{Dipartimento di Fisica ``Enrico Fermi'',
Universit\`a di Pisa, largo Pontecorvo 3, Pisa I-56127 Italy
\and
  INFN,
 Sezione di Pisa, Largo B. Pontecorvo 3, I-56127, Italy}

   \offprints{G. Valle, valle@df.unipi.it}

   \date{Received 20/07/2012; accepted 26/10/2012}

  \abstract
   {   Theoretical stellar evolutionary models are still affected by not
negligible uncertainties due to the errors in the adopted physical
inputs.  
}
   {  In this paper, using our updated stellar evolutionary code, we
quantitatively evaluate the effects of the uncertainties in the main
physical inputs on the evolutionary characteristics of low mass
stars, and thus of old stellar clusters, from the main sequence to the
zero age horizontal branch (ZAHB). 
To this aim we calculated more than 3000 stellar tracks and
isochrones, with updated solar mixture, by changing the following
physical inputs within their current range of uncertainty:
$^{1}$H(p,$\nu e^+$)$^{2}$H, $^{14}$N$(p,\gamma)^{15}$O, and
triple-$\alpha$ reaction rates, radiative and conductive opacities,
neutrino energy losses, and microscopic diffusion velocities.  
}
{   The analysis was conducted performing a systematic variation on a
fixed grid, in a way to obtain a full crossing of the perturbed input
values.  The effect of the variations of the chosen physical inputs on
relevant stellar evolutionary features, such as the turn-off luminosity, the
central hydrogen exhaustion time, the red-giant branch tip luminosity, the
helium core 
mass, and the ZAHB
luminosity in the RR Lyrae region are analyzed in a statistical way.
}
  {  We find that, for a 0.9 $M_{\sun}$ model, the cumulative uncertainty
  on the turn-off, the red-giant 
branch  tip, and the ZAHB   
luminosities accounts for $\pm 0.02$ dex, $\pm 0.03$ dex, and $\pm 0.045$ dex
respectively, while the  central hydrogen  
exhaustion time varies of about $\pm 0.7$ Gyr. 
For all examined features the most relevant effect is due to
the radiative opacities uncertainty; for the later evolutionary stages
the second most important effect is due to the triple-$\alpha$ reaction
rate uncertainty.
For an isochrone of 12 Gyr, we find that the isochrone turn-off log
luminosity   
varies of $\pm 0.013$ dex, the mass at the isochrone turn-off
varies of $\pm 0.015$ $M_{\sun}$, and the difference between ZAHB and turn-off
log-luminosity varies of $\pm 0.05$ dex.
The effect of the physical uncertainty affecting the age inferred from 
turn-off luminosity and from the vertical method are
of $\pm$ 0.375 Gyr and $\pm$ 1.25 Gyr respectively.

 }
{}


   \maketitle

\section{Introduction}\label{sec:intro}

An evaluation of the global uncertainty in stellar models due to 
the micro-physical inputs is essential for understanding the actual
significance  
of the application of these models when deriving quantitatively fundamental
stellar, and even 
 cosmological, parameters. In fact,
theoretical models for the structure and evolution of stars are indispensable
tools in many 
 astronomical research areas. Much fundamental information on resolved stellar
 populations  
 otherwise  inaccessible, as for example the age, is obtained by comparing
 observational 
  data and theoretical predictions. Furthermore, evolutionary models
play a crucial role also in  
  the studies of unresolved stellar populations, since they are a fundamental 
  ingredient for the required stellar population synthesis tools.  
  As such, in the last decades a huge effort has been focused 
 on refining the accuracy and reliability of evolutionary predictions. As a
 result, 
 stellar evolution theory
has become one of the most robust area of astrophysical research allowing a
 firm understanding  
 of the main stellar population characteristics. 

However, the current generation of stellar 
models are still affected by not negligible uncertainties, as proved by the 
discrepancies among the stellar tracks and isochrones computed by different
authors adopting  
different input physics and/or prescriptions for the treatment of processes
occurring in stars.  
These models are in fact the result of complex calculations relying on many
physical assumptions (i.e. equation of state,  
radiative and conductive opacities, nuclear reaction cross sections, neutrino
emission rates, etc.),  
algorithms describing physical processes (i.e. convective transport,
rotation, etc.), and input parameters  
(i.e. initial metallicity, initial helium abundance, heavy-element mixture,
etc.), each affected by uncertainties.  

A first estimate of the uncertainty in stellar models can be obtained by
comparing 
the results provided by different  
authors and codes \citep[see e.g.][]{padova09,dotter07,teramo06}. 
However, this kind of approach
does not allow to disentangle  
and quantify the contributions of the various uncertainty sources. 

A more suitable approach consists in changing a given input physics at a time
keeping all the other inputs and parameters fixed.  
For hydrogen-burning models of metal poor low-mass stars, an early example of
this technique is provided by \citet{chaboyer1995}  and  
subsequently extended to more advanced evolutionary phases by
various authors such as
\citet{Cassisi1998,brocato1998,Castellani1999}, and for    
white dwarf cooling models by \citet{PradaMoroni2002}.  
Many other papers followed this approach focusing on different mass ranges
and/or evolutionary phases  
\citep[see
  e.g.][]{Castellani2000,Imbriani2001,
  Salaris2002b,Imbriani2004,Weiss2005,PradaMoroni2007,cefeidi,Tognelli2011}.

However, the previous method does not allow to quantify the possible
interactions among the different input physics.  
A more systematic and exhaustive analysis consists in varying simultaneously
all the main 
input physics adopting either a Monte Carlo technique 
 \citep[see e.g.][]{Chaboyer1996,Chaboyer1998, Chaboyer2002, 
 Krauss2003, Bjork2006} or a systematic variation of the inputs in a fixed
 grid. These kind of studies are clearly much more time  
 consuming than the previous ones, since they require the computation of a
 huge number of stellar models. This is the  
 reason why a thorough analysis of this kind is still lacking. 
 
In the present paper we begin to fill this gap focusing only on the impact on
stellar models of low-mass stars, from the 
main sequence (MS) to the zero age horizontal branch (ZAHB), of the
uncertainties affecting  
some of main physical inputs adopted in modern stellar evolution codes.

The aim is twofold: first, to give an estimate of the cumulative
effect of physical uncertainties on the tracks, isochrones
and on the main stellar quantities; second, to break down such 
a global variability, evidencing the effects of the single physical 
inputs on different stages of stellar evolution.

Relying on the computation of a large number of stellar models 
(i.e. 3159 stellar tracks and 567 isochrones), we performed
a rigorous statistical analysis of the effects of the variations
of the chosen 
 physical inputs on relevant stellar evolutionary features, such as the
 turn-off luminosity, the 
central hydrogen exhaustion time, the luminosity and helium-core mass at the
 Red Giant Branch (RGB) tip, and the ZAHB luminosity in the RR Lyrae region
at $\log T_{\rm eff}$ = 3.83.

Section \ref{sec:metodo} is devoted to a description of the method employed
for the calculations and of the physical inputs
relevant to the calculations and of their current uncertainty. 
Section \ref{sec:track} and \ref{sec:statistic_track} present and analyze the
results on the global uncertainty of 
our reference model due to all the variations of the chosen inputs within
their estimated uncertainty. In
Sec. \ref{sec:iso} and \ref{sec:statistic_isoc} we report the corresponding
uncertainty on the isochrones. 
Some concluding remarks are given in  Sect. \ref{sec:conclusions}.

\section{Description of the method}\label{sec:metodo}

For the present analysis we are only interested in quantifying the theoretical
uncertainty  
affecting low-mass stellar models due to the cumulative effects of the
uncertainties in the  
main input physics. We focus on the typical member of an old
stellar 
population, choosing 
for reference a $M = 0.90$ $M_{\sun}$ model with initial 
metallicity $Z = 0.006$ -- suitable for metal-rich galactic globular clusters 
and in the Magellanic Clouds -- with heavy elements solar
mixture by \citet{AGSS09}.  
The initial helium abundance, $Y = 0.26$,  was obtained 
following the often adopted linear helium-to-metal enrichment law given by:
$Y=Y_p+\frac{\Delta Y}{\Delta  Z}Z$, with cosmological $^4$He abundance
$Y_p=0.2485$
\citep{cyburt04,steigman06,peimbert07a,peimbert07b}.
In this work we assume 
$\Delta Y$/$\Delta Z=2$, 
a typical value for this quantity, still affected by several
important sources of  
uncertainty \citep{pagel98,jimenez03,flynn04,gennaro10}.

The adopted stellar evolutionary code, FRANEC, has been extensively described 
in previous papers \citep[and references therein]{cariulo04,scilla2008}. 
A detailed discussion of the recent updates of the physical inputs can be
found in \citet{cefeidi} and 
\citet{Tognelli2011}. The code adopted here is the same used for the
construction of the Pisa Stellar Evolution Data 
Base\footnote{\url{http://astro.df.unipi.it/stellar-models/}} for low-mass
stars, 
as illustrated in  
\citet{database2012}, where a detailed description of the inputs of
the stellar evolutionary code and of the ZAHB\footnote{We followed a synthetic
  method to build the ZAHB models as we did not evolve the tracks 
 through the He-flash. The ZAHB models have been computed by accreting  
envelopes of different mass extensions onto the He-core left at the tip of the
RGB.} construction technique can be 
found.    

The mixing length formalism \citep{bohmvitense58} is used to treat the
convective transport 
in superadiabatic regions. In such a scheme the convection
efficiency depends on the mixing length parameter $l= \alpha_{\rm ml} H_p$, where 
$H_p$ is the local pressure scale height and $\alpha_{\rm ml}$ a free
parameter to be calibrated.  

The treatment of the superadiabatic convective transport is one of the weakest
aspect  
in current generation of stellar evolution codes and consequently one of the
main sources  
of theoretical uncertainty, which, however, mainly affects effective
temperature predictions while luminosity evaluations are affected in very much
less extent.
Such an uncertainty should be considered as  
systematic and due to an oversimplified treatment of a very complex
phenomenon, whose precise physical description is not yet available.
Since we are  
interested only in evaluating the cumulative effect of the main input physics
uncertainties, we  
did not take into account the effect of the large error in the efficiency of
the superadiabatic  
convective transport, i.e. on the $\alpha_{\rm ml}$ value. For this reason,
all the models  
computed in the present analysis adopt $\alpha_{\rm ml}=1.90$.
However, a change of the $\alpha_{\rm ml}$ value chosen in the computations 
might in principle affect the estimate of the extent of the cumulative
uncertainties in 
 stellar predictions. For this reason, as shown in detail in Appendix
 \ref{app:mixlen},  
 we computed additional sets of models with two different mixing-length
 parameter values,  
 namely $\alpha_{\rm ml}=1.70$ and 1.80, checking the robustness of the results 
 presented in the paper for a reasonable change in the mixing-length value.

\subsection{Input physics uncertainties}

The computation of stellar models
relies on the detailed  
knowledge of matter and radiation behavior in the temperature and density
regimes typical of stellar interiors.  
These input physics include: radiative and conductive opacities,
equation-of-state (EOS),  
nuclear reaction cross sections, neutrino emission rates (i.e. plasma, photo,
pair, and bremsstrahlung processes).

In this subsection, we briefly summarize the uncertainty on the
up to date input physics, focusing only on those which play a relevant role
in the structure and evolution of  
the reference case, with the notable
exception of the EOS.  
In fact, in spite of its importance in modeling stellar interiors, a
detailed and rigorous evaluation  
of the propagation of the EOS uncertainties into the final model is a
difficult task.  
The reason is that the main thermodynamic quantities required for computing
stellar models  
(i.e. pressure, temperature, density, specific heat, adiabatic gradient, etc.) 
are available only in the form of numeric tables, where the values of the
various quantities are  
given without the associated uncertainty. Furthermore, those thermodynamic
quantities are related  
in a not trivial way. Thus, we preferred to fix the EOS as it
was without errors  
rather than adopting a too crude treatment.  

Present calculations used the most recent version of the OPAL
EOS\footnote{Tables available at  
  \url{http://rdc.llnl.gov/EOS_2005/}}, namely the 2006 release 
\citep{rogers1996, rogers02}.

Anyway an estimate of the impact of  two different choices of EOS is
  presented in Appendix~\ref{app:eos}, where we evaluate the differences in
  the evolutionary features computed with OPAL and FreeEOS \citep{irwin04}.

\paragraph{Nuclear reaction rates}
  
Regarding the nuclear reactions relevant for hydrogen and helium burning
phases,  
we adopted the same cross sections detailed in \citet{database2012}. 
In order to avoid the huge and useless explosion of computational time that
would  
result taking into account the uncertainty in the cross sections of all the
 nuclear reactions belonging to the hydrogen and helium burning networks, 
 we limited the analysis only  to the three main reactions, namely  
 the $^{1}$H(p,$\nu e^+$)$^{2}$H, $^{14}$N$(p,\gamma)^{15}$O, and triple-$\alpha$.
In fact the $^{1}$H(p,$\nu e^+$)$^{2}$H and $^{14}$N$(p,\gamma)^{15}$O are the
lowest cross sections which thus drive, respectively, the efficiency 
of the proton-proton chain and the CNO cycle, while the triple-$\alpha$ cross
section influences both the core helium flash and the ZAHB luminosity.

Experimental measurements of the $^{1}$H(p,$\nu e^+$)$^{2}$H cross 
section are not available, thus one has to rely only on theoretical 
models. The uncertainty in the S$_{\rm pp}$(0) factor is of the order of 1\%
 \citep{kamionkowski1994, adel98, adel2011}. However, the current version of 
 the FRANEC, as many other evolutionary codes, calculates the
  $^{1}$H(p,$\nu e^+$)$^{2}$H reaction rate following the analytical
 approximation provided by the NACRE compilation \citep{nacre}, whose 
accuracy with respect to the tabulated rates is better than 3\%. 
 This is the value of uncertainty we adopted as it is the dominant source of 
 error in the rate calculations.

For  $^{14}$N$(p,\gamma)^{15}$O reaction rate we used the recent estimate
by \citet{14n}. The uncertainty on the analytical fit given in that
paper is about 8$\div$10\% in 
the range of temperature [$10^6$ - $10^8$] K.  In the
calculations we assumed an uncertainty of 10\%.

\citet{Fynbo2005} reported new measurements for the triple-$\alpha$ rate:
however, in the temperature range of interest, the differences with
respect to the most widely adopted rate (NACRE,
\citealt{nacre}) are within the uncertainty evaluated in the NACRE compilation, 
reaching a maximum of $\approx$ 20\% at temperatures of about $10^8$ K
\citep[see e.g.][]{Weiss2005}.   
For this reason we decided to take the error quoted by NACRE (20\%), in the
temperature range of interest, as an estimate of the uncertainty on
the triple-$\alpha$ reaction rate.

\paragraph{Radiative opacity}

The radiative opacity is one the most important input for stellar
model calculations. 
As described in \citet{database2012}, we implemented in our code the
opacity tables
provided in 2005 by OPAL\footnote{The OPAL radiative opacity can be
found at the URL: \url{http://opalopacity.llnl.gov/new.html}}   
\citep[see e.g. ][]{iglesias96} for $\log T({\rm K}) > 4.5$ and by
\citet{ferg05} 
for lower temperatures. Both opacity tables have been computed adopting
the solar heavy-element mixture by \citet{AGSS09}. 
   
In spite of the crucial importance of radiative opacity in the computation of
stellar models,  
the quoted tables do not contain the uncertainty associated to the single
Rosseland mean opacity  
coefficients. Furthermore, neither \citet{iglesias96} nor \citet{ferg05}
provide even a rough estimate  
of the accuracy of the results. The same occurs for the radiative opacity
tables made available  
by the Opacity Project \citep[hereafter OP, ][]{OP2004, OP2005}. 
{Given such a situation, a first idea of the accuracy level of the current
  radiative opacity 
generations can be achieved by comparing the results provided by different and
independent groups.  
\citet{rose2001} performed this kind of analysis for a plasma mixture very
similar in composition, temperature, and density  
to that in the Sun's center finding an agreement between 
Rosseland mean opacity coefficients
provided by  different codes   
at the level of 5\%. The agreement gets worse at different plasma conditions,
as shown by comparing 
 the results provided by OPAL and OP (hereafter $k_{\rm r}^{\rm OPAL}$ and
 $k_{\rm r}^{\rm OP}$) over their whole  
 validity domain, which are in agreement within about 10-15\%  \citep{OP2004,
   OP2005}. } 

\begin{figure*}
\centering
\includegraphics[width=6.3cm,angle=-90]{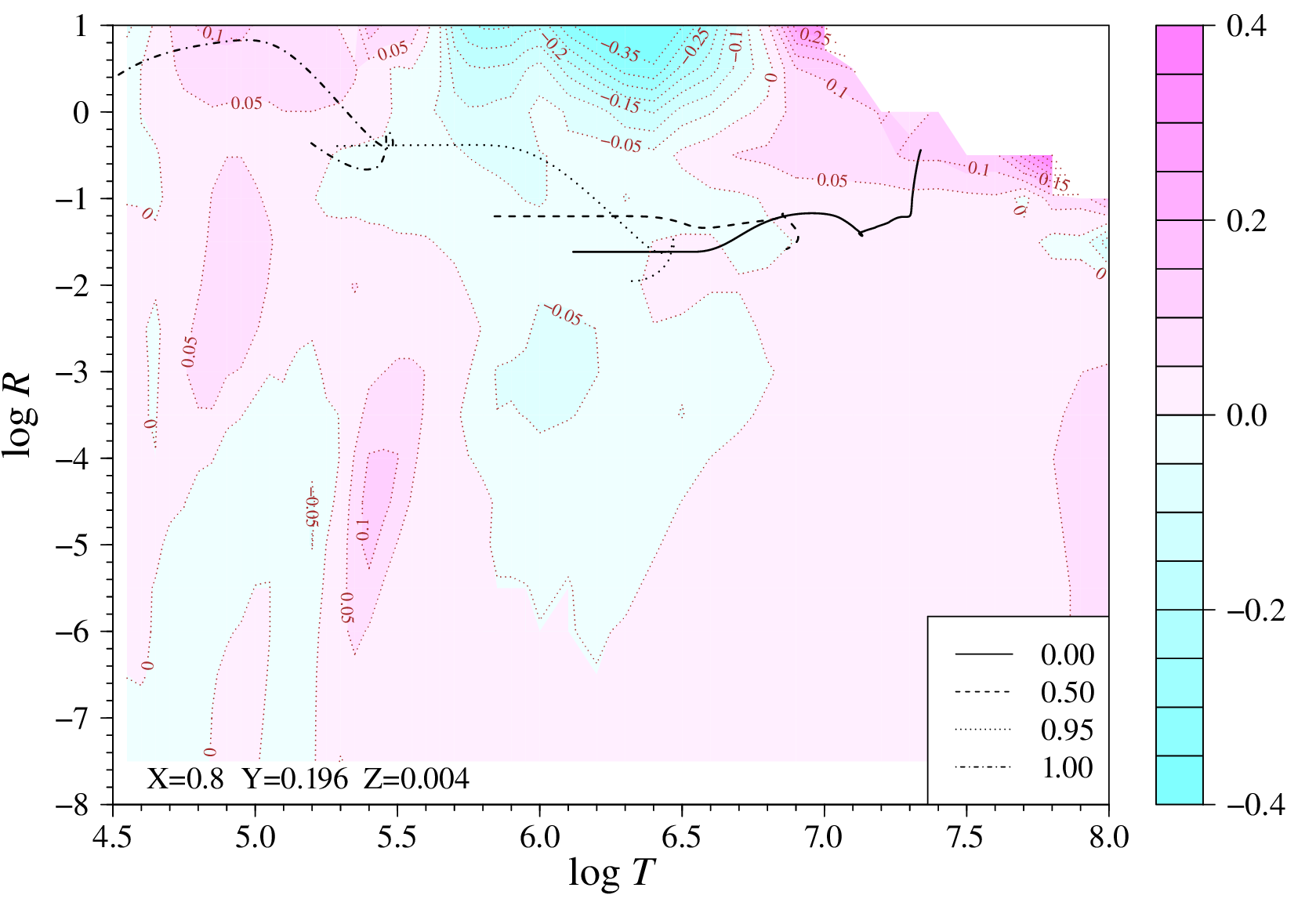}\hspace{0mm} 
\includegraphics[width=6.3cm,angle=-90]{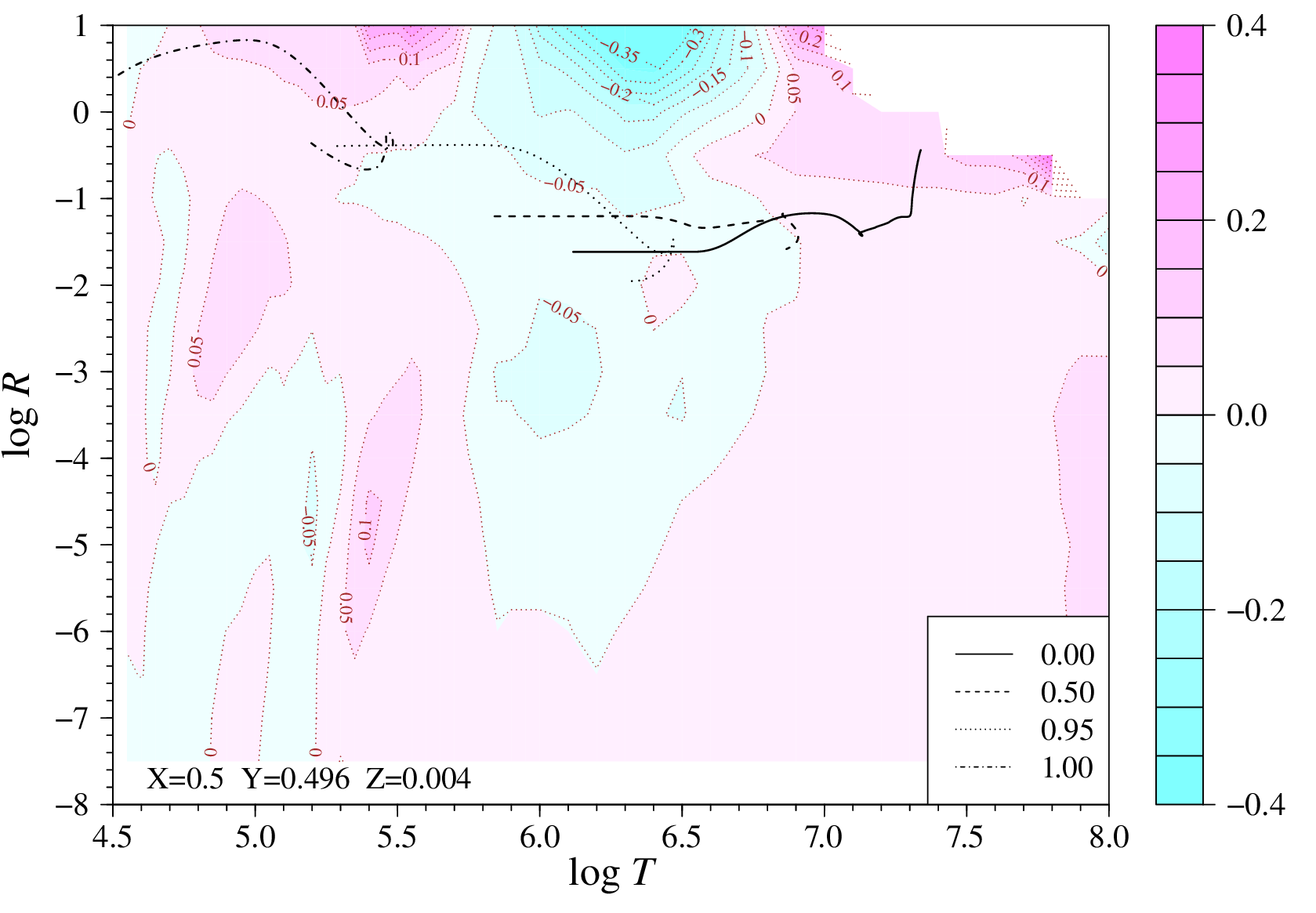}\\\vspace{2mm}
\includegraphics[width=6.3cm,angle=-90]{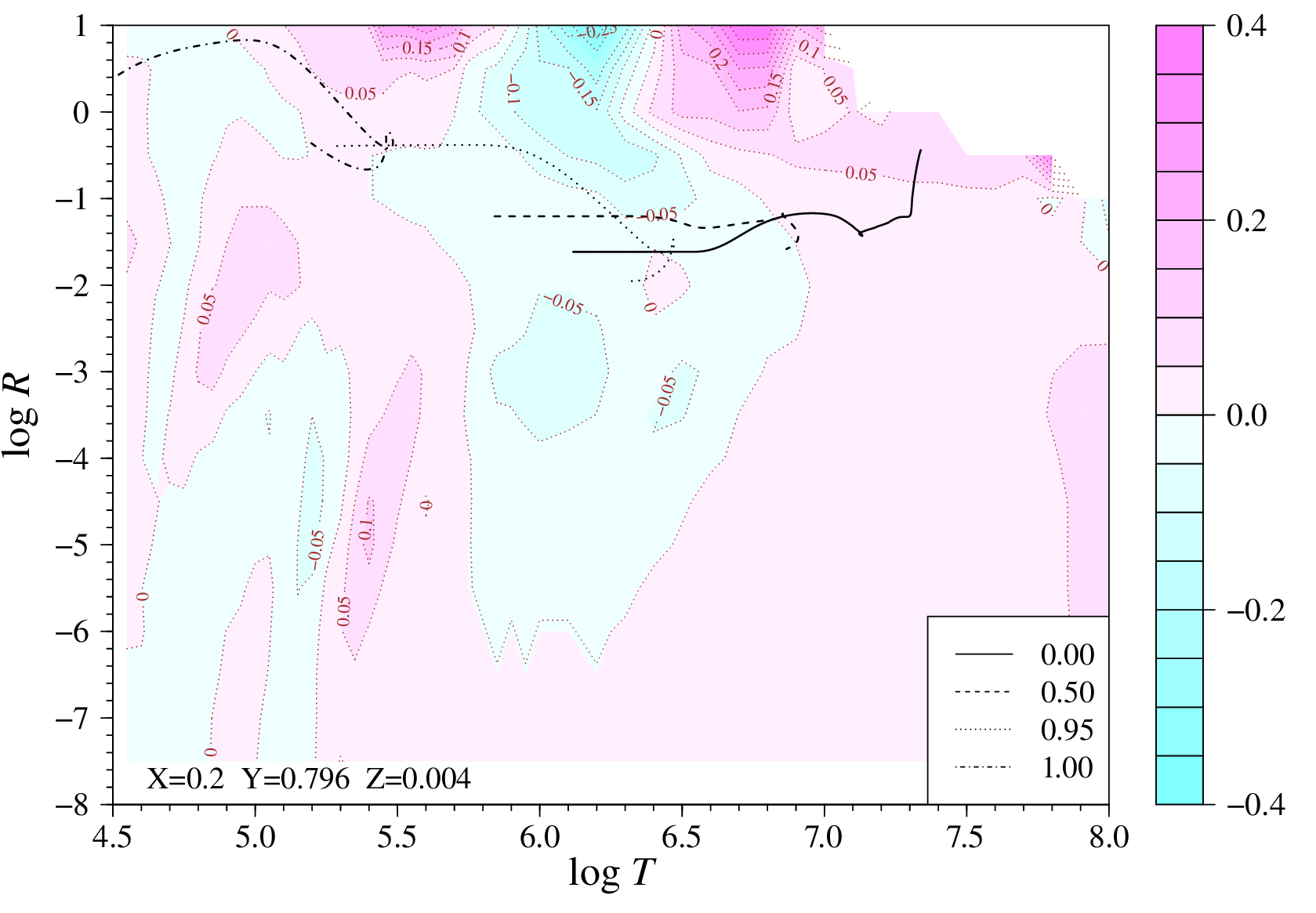}\hspace{0mm}
\includegraphics[width=6.3cm,angle=-90]{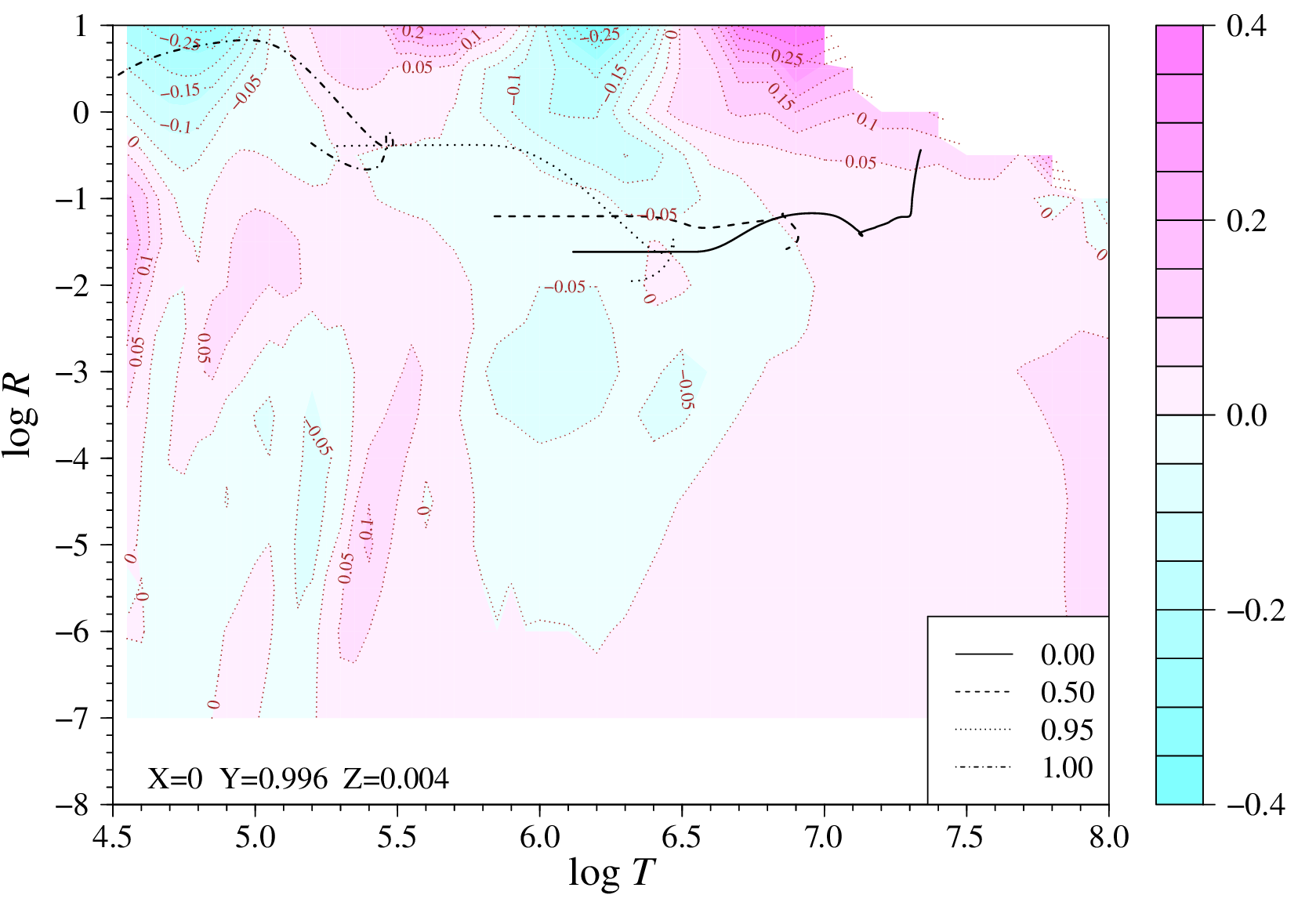}
\caption{Contour plots of the difference between OPAL and OP
calculations for different hydrogen abundance $X$ and metallicity $Z$ = 0.004. 
The temperature $T$ is in K, while $R$ is in g cm$^{-3}$ K$^{-3}$.
The
coloured scale marks the values of the relative difference $(k_{\rm r}^{\rm
  OPAL} - k_{\rm  r}^{\rm OP})/k_{\rm r}^{\rm OPAL}$. 
Solid
line: evolutionary path of the stellar center; dashed line:
evolutionary path of the
0.50 mass fraction of the structure; dotted line: path of the 0.95 mass
fraction of the structure; dot-dashed line: path of the 0.99974 mass
  fraction  of the structure, labeled as 1.00.}
\label{fig:OPALvsOP}
\end{figure*}

To further extend the investigation, we evaluated the differences
 in the 
($\log T$, $\log R$) plane between OPAL and OP radiative opacities computed for 
\citet{AGSS09} solar elements mixture. As in OPAL and OP works, we used here
the variable $R  = \rho/T_6^3$, where $\rho$ is the mass density and $T_6 =
10^{-6} \times T$ with $T$ in K. 
Figure \ref{fig:OPALvsOP} shows the values of $(k_{\rm r}^{\rm OPAL} - k_{\rm
  r}^{\rm OP})/k_{\rm r}^{\rm OPAL}$ 
in four contour plots,
for different hydrogen abundances (i.e. $X$ = 0.8, 0.5, 0.2, and 0.0) and
  metallicity $Z$ = 0.004. 
 Notice that we did not plot the comparison for the same metallicity of our
 reference track, i.e. $Z$ = 0.006,  
 because the corresponding tables are not available neither in OPAL nor OP
 calculations,  
 thus we preferred to use the nearest value of the metallicity for which both
 groups provide the opacity tables, 
  i.e. $Z$ = 0.004, rather than interpolate.   
The computed evolution of the standard stellar model with $M = 0.90$
$M_{\sun}$, $Z = 0.006$, $Y = 0.26$ is superimposed to each panel.  
To give an indication of the region spanned in the plane by the whole
stellar structure, we showed the path of four 
different mass fractions of the structure, from the center
to the outer layers, corresponding to the 0.99974 mass fraction (respectively,
labels from 0.00 to 1.00 in Fig. 
\ref{fig:OPALvsOP}). 

Looking at the panels of Fig. \ref{fig:OPALvsOP} we see that, for
the selected model, a 5\% uncertainty on the values of
radiative opacities is adequate, so we adopted this value in the calculations. 
We assumed the same uncertainty also for the low-temperature 
radiative opacity coefficients computed by \citet{ferg05}.

\paragraph{Conductive opacity}

In regions of stellar interiors characterized by electron degeneracy, 
as in the helium core of low-mass red giant stars,
thermal electron conduction, elsewhere inefficient, contributes significantly
to the 
energy transport. Thus, in these regimes it becomes an important input
 which affects  the stellar model. 

For the electron conduction opacities $k_{\rm c}$, we adopted the 
results by  
\citet{casspote07}, which improved the conductive opacities by \citet{pote99a}.
In the mildly degenerate, weakly or moderately coupled
plasmas, typical of the red giant helium cores, the dynamical plasma
screening may be non-negligible. This effect was studied by \citet{Lampe1968b},
but it was not included in either \citet{hubbard69} 
or in \citet{casspote07} calculations.
Using an estimate for the correction due to the dynamical plasma
screening,
Potekhin suggested (2011, private communication) that \citet{casspote07}
opacities are uncertain by 5\% at the center and few ten
percent in the 
outer  
regions of the helium core.
However, in our case the uncertainty is lower than the 10\% in about the 80\%  
of the He-core mass, which roughly corresponds to the zone dominated by 
electron conductivity. So we adopted a 5\% uncertainty as sensible choice
for conductive opacity.

\paragraph{Neutrino emission rates}

At high temperature and density, several processes of
neutrino emission  
are efficient (i.e. plasma, photo, pair, and bremsstrahlung processes), which
provide additional and  
relevant energy loss channels in stellar interiors, with the consequent
significant effect on  
stellar structures. During the red giant phase for the chosen reference
case, in the dense and not extremely hot helium core,  
neutrino losses are dominated by the plasma neutrino emission process, whose
efficiency  
affects important quantities of red giant evolution, such as the helium
core mass and the stellar luminosity  
at the helium-flash onset. 
 
We followed the fitting formulae given in \citep{haft94} to compute the 
plasma neutrino emission rate. 
As reported in literature \citep{itoh96, haft94}, the  accuracy of that
approximation 
is better than 4\% for almost every value of temperature and density and
better than 5\% everywhere.  In the calculations, we adopted an uncertainty of
4\%.

\paragraph{Helium and heavy elements diffusion velocities}

The computation of the diffusion velocities of helium and heavy elements 
in our code is performed by the routine developed by \citet{thoul94}
\citep[for more details see 
e.g.][]{Castellani1997,database2012}. These diffusion velocities are generally
thought
to be accurate at 10$\div$15\% \citep[see e.g.][]{thoul94}, thus, in the
present computations we adopted  
a conservative uncertainty of 15\%. 
 
The selected physical inputs and their assumed uncertainty are listed in Table
\ref{table:inputfisici}. 

\subsection{Method to evaluate the cumulative uncertainties}

The complexity of stellar evolution calculations hampers an analytical 
evaluation of   
the impact of the variation of the chosen inputs on stellar models
calculation, so that the problem must be addressed by direct computation of
perturbed stellar models, i.e. models adopting physical inputs perturbed
within the uncertainties.  

Several choices are available for the design of the sample to
investigate. 
As already discussed in the introduction, the widely followed approach is to
allow for the variation of one parameter at a time, making possible the
quantification 
of the separate effect of the studied inputs. This approach is however
vulnerable 
in presence of interactions among the inputs, since it does not allow to
detect a possible synergyc effect due to
the variation of several inputs at a time. 
 
To avoid this weakness, we employ a more robust, even if
much more  
computationally expensive, technique namely a systematic variation of the
inputs on a 
fixed grid. For each physical input, we introduced a three-values
multiplier $p_i$ with value 
1.00 for the reference case and values $1.00 \pm \Delta p_i$ for perturbed
cases ($\Delta p_i$ is the uncertainty listed in Table
\ref{table:inputfisici}), which defines 
the range of variations. For each stellar 
track calculation, a set of multiplier values (i.e. $p_1, \ldots, p_7$
for the seven input physics  
allowed to vary) is chosen and kept constant during the evolution of the
stellar structure. In order to cover the whole parameters space, 
calculations of stellar tracks were performed for a full crossing, i.e. each
parameter value $p_i$ was crossed with all the values of the other parameters
$p_j$, with $j \neq i$.   
In this way, we computed stellar models for all the possible sets of
multiplier values. 
A total of $3^7 = 2187$ tracks were then computed, with same mass, chemical
composition and $\alpha_{\rm ml}$. 

The technique allows the exploration of the edge of the variability region,
and it
is robust in presence of interaction among the inputs of the
calculations, 
since it does not assume physical independence of the individual processes.

An alternative approach to the problem would require the use of Monte Carlo
simulations \citep[see e.g.][]{Chaboyer1996,Chaboyer1998, Chaboyer2002,
 Krauss2003, Bjork2006}: this technique 
allows the construction of probabilistic confidence 
intervals for the most interesting evolutionary features. 

The grid technique employed is distribution-free, i.e. it needs only the
specification of a sensible range of 
variation for each physical input, and it does not rely on explicit
specification of the parent distributions of the physical parameters that we
varied. 
The simpler grid variation technique has the
additional advantage to be less computationally expensive
than a Monte Carlo. In fact a key value to evaluate for a Monte Carlo
simulation is the number of runs $N$ to perform in order to have an acceptable
coverage 
of the input parameter hyperspace. As an example, dividing the
range 
of variation of all the parameters in 4 zones according to quartiles, the
hyperspace of parameters will consist on $n = 4^7 = 16384 $ hypercells.  
In the hypothesis that all the parameters have uniform distribution,
the probability of having $k$ hit of a cell during the Monte Carlo simulations
is given by Poisson density function with mean $N/n$: $P(k, N/n) = k^{N/n} /
(N/n)! \; {\rm e}^{-k}$. 
If we require a probability $\alpha$ that a cell has zero hit -- i.e. it is
unexplored during the simulations -- we have: $P(0, N/n) = \alpha$.
Solving for $\alpha = 0.15$, an acceptable compromise between coverage of the
hyperspace and
accuracy of the output, it results $N \approx 31000$ run, which is an order of
magnitude bigger than our sample size.

\begin{table}[ht]
\centering
\caption{Physical inputs perturbed in the calculations and their assumed
  uncertainty. In parentheses are defined the abbreviations used in the
    following Tables.} 
\label{table:inputfisici}
\centering
\begin{tabular}{llc}
  \hline\hline
 description & parameter & uncertainty \\
 \hline
  $^{1}$H(p,$\nu e^+$)$^{2}$H reaction rate (pp)      & $p_1$ & 3\%  \\
  $^{14}$N(p,$\gamma$)$^{15}$O reaction rate ($^{14}$N)     & $p_2$ & 10\% \\
  radiative opacity ($k_{\rm r}$)                   & $p_3$ & 5\%  \\
  microscopic diffusion velocities ($v_{\rm d}$)             & $p_4$ & 15\% \\
  triple-$\alpha$ reaction rate  (3$\alpha$)               & $p_5$ & 20\% \\
  neutrino emission rate  ($\nu$)                      & $p_6$ & 4\%  \\
  conductive opacity ($k_{\rm c}$)                  & $p_7$ & 5\%  \\
   \hline
\end{tabular}
\end{table}

\section{Global physical uncertainty in stellar models}
\label{sec:track}

The availability of a large set of stellar models covering all the possible 
combinations of simultaneously perturbed input physics allow us to 
quantify the cumulative physical uncertainty in stellar models. 
 
\begin{figure*}
\centering
\includegraphics[width=6.3cm,angle=-90]{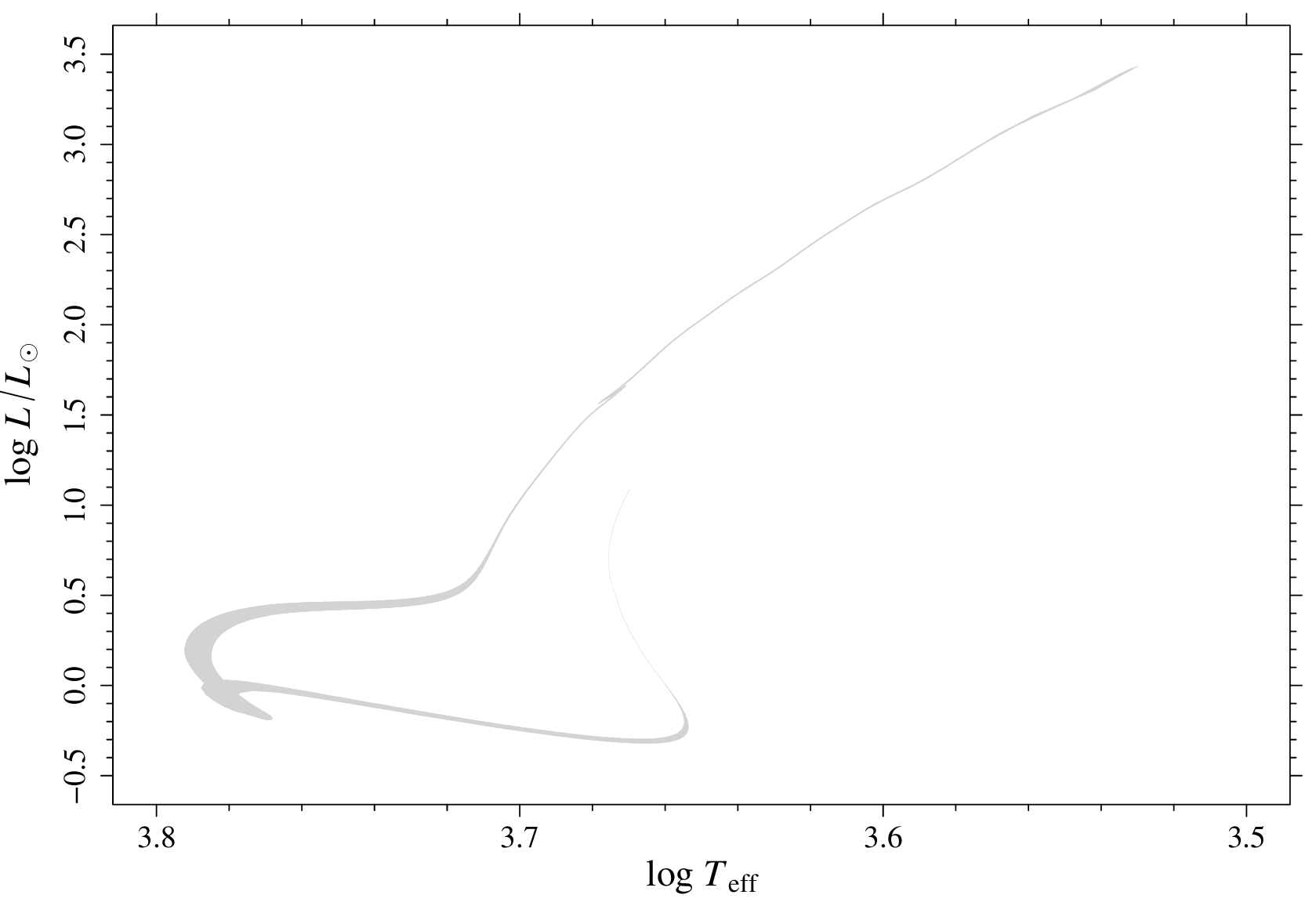}
\includegraphics[width=6.3cm,angle=-90]{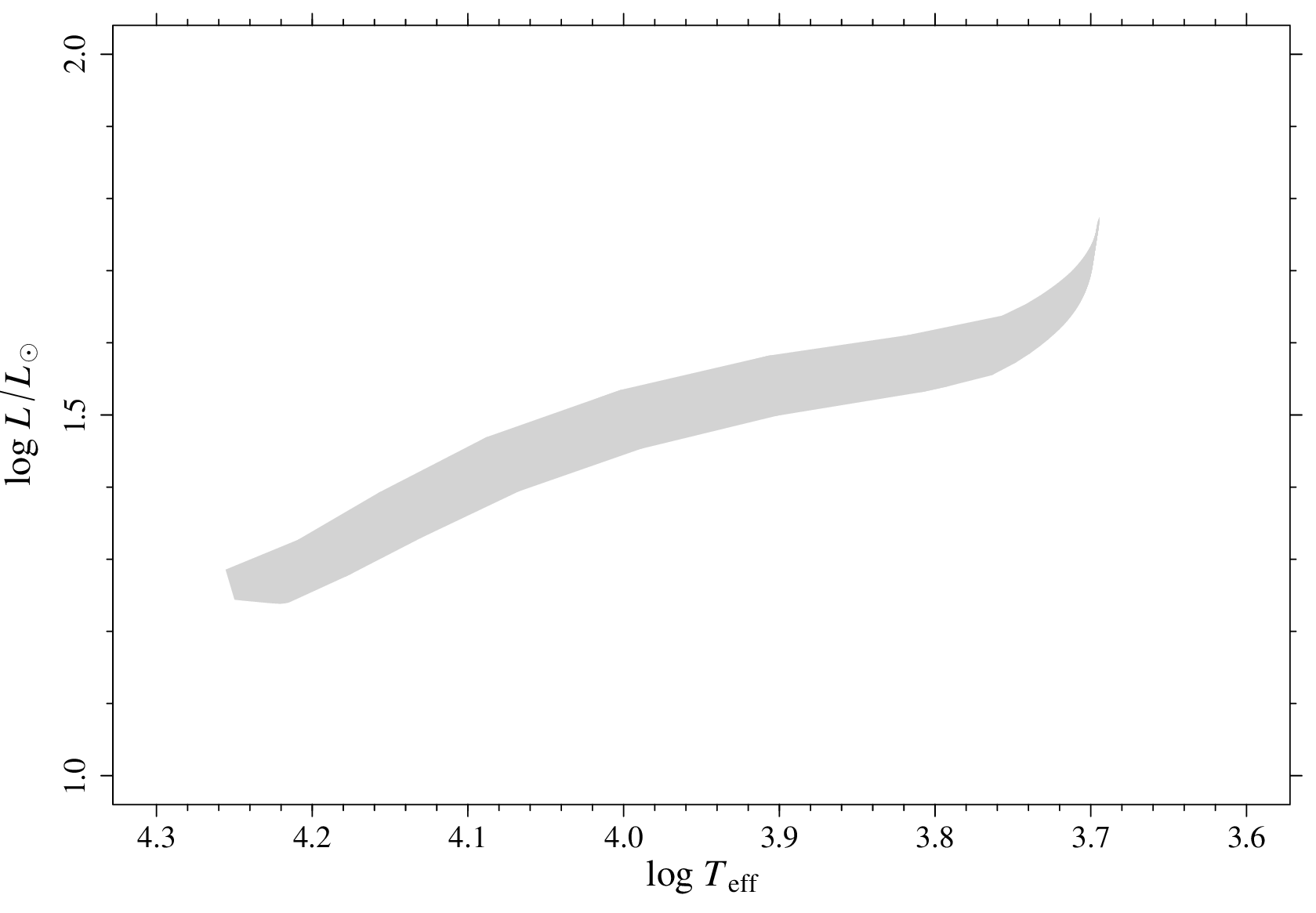}
\caption{Left panel: Hertzsprung-Russell (HR) diagram showing the error stripe
  due to 
 the variation of all the seven analyzed physical inputs (i.e. $p_1, \ldots,
 p_7$ in 
 Table \ref{table:inputfisici})
 on the stellar track with $M$ = 0.9 
  $M_{\sun}$, $Z$ = 0.006, $Y$ = 0.26 from pre-main sequence to helium flash.
The narrowing of the error stripe in the RGB is 
due to the fact that the perturbed stellar models are disposed along the
tracks itself. See text for details.
Right panel: as in the left panel, but for the ZAHB.}
\label{fig:nastro-09}
\end{figure*}

The combined effect in the theoretical plane ($\log T_{\rm eff}$, $\log
L/L_{\sun}$) of the variation of all the seven physical inputs (i.e. $p_1,
\ldots, p_7$ in Table~\ref{table:inputfisici}) 
 selected for the calculations is displayed in Fig.~\ref{fig:nastro-09},  
where the track computed with the standard inputs is enveloped by an error
stripe constructed by considering the region of the plane spanned by the
perturbed stellar models. To the best of our knowledge, this is the first time  
that an error stripe is computed and plotted for theoretical stellar tracks. 
A detailed description of the technique employed for the 
construction of the stripe is given in Appendix \ref{app:nastro}.

\begin{figure*}
\centering
\includegraphics[width=6.3cm,angle=-90]{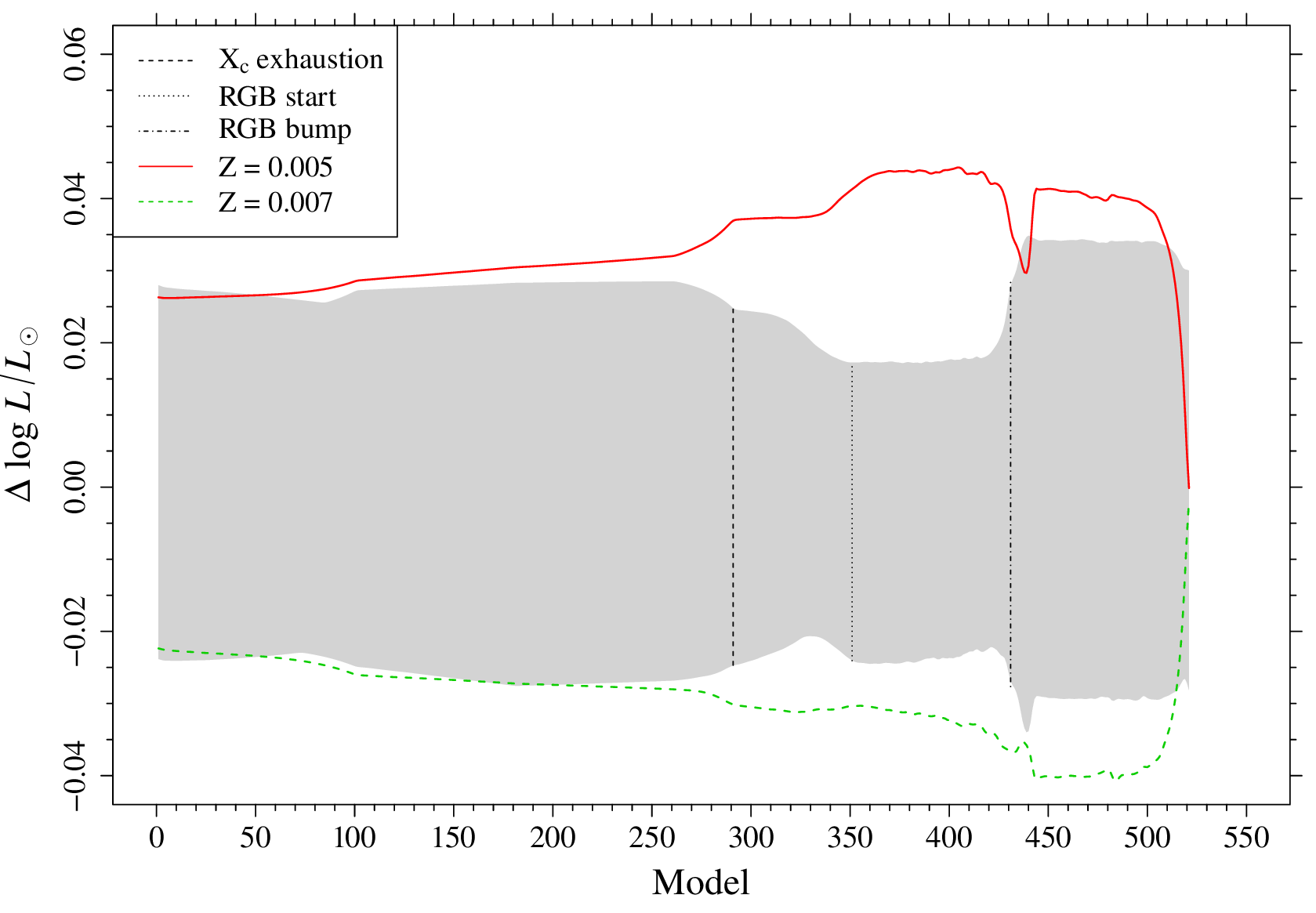} 
\includegraphics[width=6.3cm,angle=-90]{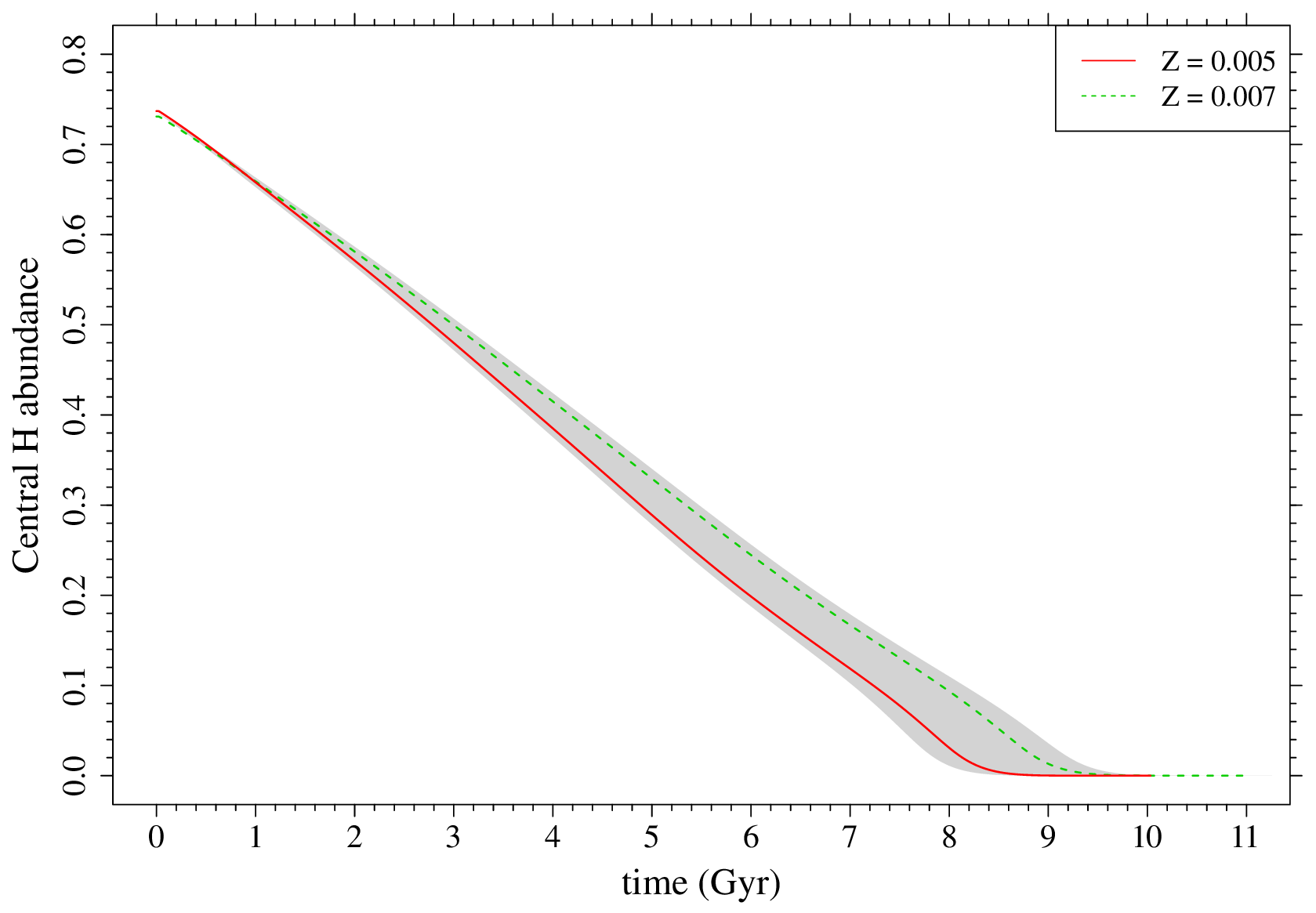} 
\caption{Left panel: $\Delta \log L/L_{\sun}$ with respect to the track
  computed with 
standard inputs from ZAMS to helium flash. The comparison was performed on a set
of reduced tracks with the same number of homologous points. Right panel:
central hydrogen abundance $X_{\rm c}$ vs. time (Gyr). In both panels, the red
solid and green 
dashed lines show for comparison the output of models with standard physical
inputs and $Z$ = 0.005, 0.007 respectively.} 
\label{fig:variazioni}
\end{figure*}

The considerable narrowing of the error stripe in the RGB is 
due to the fact that the perturbed stellar models are disposed along the
tracks itself and do not imply the vanishing of the uncertainty. 
This effect is best evidenced in left panel of Fig.~\ref{fig:variazioni} which
displays the range in $\Delta \log L/L_{\sun}$, computed  with respect to 
the standard track, from ZAMS to helium flash. To perform the comparison, the
raw tracks 
were reduced to a set of tracks with the same number of homologous points.
Details about the reduction procedure are reported in Appendix
\ref{app:nastro}.
We note that the error stripe has a nearly constant width of about 0.05 dex
until the central hydrogen exhaustion, while the error grows to about 0.07
dex in the final 
part of RGB, from the RGB bump to helium flash.

In order to compare this uncertainty with the one due to a variation 
in the initial chemical composition,
Fig.~\ref{fig:variazioni} also shows  
stellar models computed with standard physical input but different $Z$ and
$Y$. We  
trace a variation in [Fe/H] of $\pm 0.07$ -- a typical value of uncertainty
for a cluster of metallicity similar to our reference case --
with unchanged $\Delta Y/\Delta Z = 
2$, resulting in two sets: the first one with $Z$ = 0.005, $Y$ = 0.258 and
the second one with $Z$ = 0.007, $Y$ = 0.262. As one can see in the Figure, 
the impact of the quoted chemical variation on the predicted stellar
luminosity is essentially  
the same of the input physics uncertainties until the central hydrogen
exhaustion. At later evolutionary stages the former effect prevails, becoming
the dominant by approximately 20\% after RGB bump. 

As for evolutionary time, in the right panel of Fig.~\ref{fig:variazioni}
we show the evolution of the central hydrogen abundance X$_{\rm c}$ as a
function  
of time (Gyr). In
this case the effect of the variation of the physical inputs is larger
than the one due to the quoted changes in chemical abundances. For instance,
the range 
for the X$_{\rm c}$ exhaustion time due to physics variation is [9.83 - 11.26]
Gyr, while the one due to chemical variation is [10.03  - 11.00]
Gyr.

\begin{figure}
\centering
\includegraphics[width=6.2cm,angle=-90]{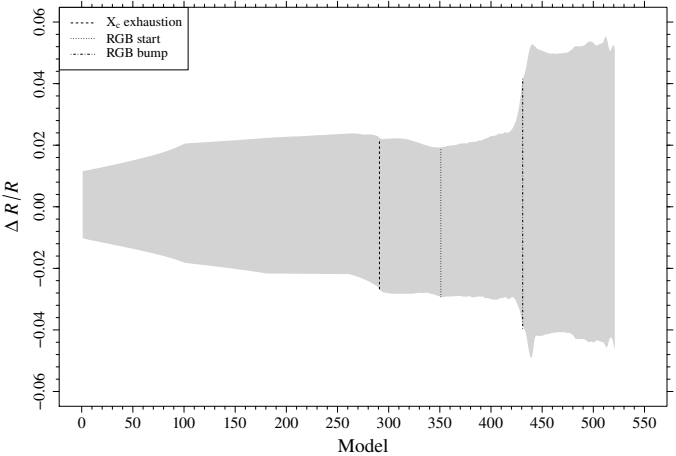}
\caption{$\Delta R/R$ with respect to the track computed with
standard inputs from ZAMS to helium flash.}
\label{fig:variazioniR}
\end{figure}

The radius is another 
quantity worth to be discussed, whose accurate
determination  
is important also in determining the properties of any orbiting exoplanet.
Regarding the impact of the perturbed physical inputs on the predicted stellar
radius, 
in Fig.~\ref{fig:variazioniR} we show the range in $\Delta R/R$, computed  with
respect to the standard track,
from ZAMS to helium flash. 
The error stripe has an increasing width from about 0.02 at the
ZAMS to about 0.04 at the
central hydrogen exhaustion, while the width grows to about 0.09 in the final
part of RGB.

We quantified also the global physical uncertainty in the theoretical 
predictions of various stellar quantities of common interest, namely  
the turn-off luminosity, the central hydrogen exhaustion time
$t_{\rm H}$, the luminosity $L_{\rm tip}$ and the helium core mass $M_{\rm
  c}^{\rm He}$  
at the RGB tip, and the ZAHB luminosity in the RR Lyrae region $L_{\rm HB}$.
In the case of turn-off luminosity we chose to investigate the luminosity of a 
point brighter and 100 K lower than the turn-off (hereafter BTO),
adopting a technique similar to the one proposed by \citet{Chaboyer1996} for
 isochrones in the (B-V, V) plane. This method has the advantage of reducing
 the intrinsic luminosity variation of the canonical turn-off, whose
 position is difficult to accurately identify both in observed
 Hertzsprung-Russell (HR) diagrams and
 in  
 theoretical tracks/isochrones, as a consequence of the almost vertical slope
 (i.e. large luminosity variation  at essentially the same effective
 temperature).  
 Then, a small fluctuation in the effective temperature
 determination  
 could have a large effect in the determination of the turn-off luminosity.

\begin{table}[ht]
\centering
\caption{Total range of variation and range half-width of the theoretical
  predictions for the  selected   quantities for our reference case,  
i.e. $M = 0.90$ $M_{\sun}$ with $Z = 0.006$ and $Y = 0.26$, due to input
physics uncertainties.}  
\label{table:incertezzetracce}
\centering
\begin{tabular}{lcc}
  \hline\hline
 quantity & variation range & range half-width\\
 \hline
  $\log L_{\rm BTO}$ & [0.334 - 0.376] dex & 0.021 dex\\
  $t_{\rm H}$ & [9.83 - 11.26] Gyr & 0.72 Gyr\\
  $\log L_{\rm tip}$ & [3.38 - 3.44] dex & 0.03 dex\\
  $M_{\rm c}^{\rm He}$  & [0.4796 - 0.4879] $M_{\sun}$ & 0.0042 $M_{\sun}$ \\
  $\log L_{\rm HB}$ & [1.52 - 1.61] dex & 0.045 dex\\
   \hline
\end{tabular}
\end{table}

Table \ref{table:incertezzetracce} lists the total range of variation in
predictions  
of the selected quantities for our reference stellar track due to current
input physics uncertainties.    
The turn-off log luminosity $\log L_{\rm BTO}$ varies in the range [0.334 -
  0.376] dex 
(range half-width 0.021 dex, $\approx 6\%$ of the value obtained with
unperturbed 
physical inputs). The total range of variation of the predicted
central  
hydrogen exhaustion time $t_{\rm H}$ is [9.83 - 11.26] Gyr (0.72 Gyr, $\approx
6.5\%$). 
The RGB tip $\log L_{\rm tip}$ and ZAHB $\log L_{\rm HB}$ log luminosities
vary, respectively, 
 in the ranges [3.38 - 3.44] dex (0.03 dex, $\approx 1\%$) and 
 [1.52 - 1.61] dex (0.045 dex, $\approx 3\%$). 
Finally, the helium core mass at the RGB tip $M_{\rm c}^{\rm He}$ varies in
the range  
 [0.4796 - 0.4879] $M_{\sun}$ (0.0042 $M_{\sun}$, $\approx 0.85\%$).

\section{Statistical analysis of physical uncertainty in stellar models}
\label{sec:statistic_track}

The large set of computed stellar models is suitable for an accurate statistical
analysis of the effect of the variation of the chosen physical inputs on
relevant stellar evolutionary features. Beside the quantification of the 
 whole range of uncertainty performed in the previous section, it is possible
 to disentangle the effects of the 
 different physical inputs on the above selected stellar quantities.

The dependence of the aforementioned evolutionary quantities on physical
inputs was explored by means of linear regression models. 
These models were constructed extracting the values of the chosen dependent
variable in 
study (i.e. $L_{\rm BTO}$, $t_{\rm H}$, $L_{\rm tip}$, $M_{\rm c}^{\rm He}$,
and $L_{\rm HB}$) from the computed stellar tracks and regressing it against the 
independent variables (more properly defined as predictor variables or
covariates),  
in our case the values of the parameter $p_i$. 

The regression model construction started from models linear in the
  physical 
inputs, since a priori we do not anticipate the need of higher power of the
inputs. Similarly, at first no interaction among the covariates were
considered.  
These choices were supported by the detailed a posteriori analyzes of each
regression model: 
it results in fact that the insertion of higher power of the predictors or
explicit 
interaction among them is not needed, because the aforementioned linear
models capture almost the whole variation of the data. 
These models included as covariates only the physical
inputs which can have an 
influence on the studied evolutionary feature: for  $L_{\rm BTO}$ and
$t_{\rm H}$ only the first four parameters of Table \ref{table:inputfisici},
while for later evolutionary stages all the parameters were used.
The models were fitted to the data with a
least-squares method using the software R 2.14.1 \citep{R}. Results of the
statistical analyzes were
considered statistical significant for $p$-value $< 0.05$.  

For the BTO log-luminosity $\log L_{\rm BTO}$ and $t_{\rm H}$ the regression
models were: 
\begin{equation}
\log L_{\rm BTO}, t_{\rm H}  = \beta_0 + \sum_{i=1}^4 \beta_i \; p_i 
\end{equation}
where $\beta_0, \ldots, \beta_4$ were the regression coefficients to be
estimated by the fit and 
$p_1, \ldots, p_4$ respectively the perturbation multipliers  (see Table
\ref{table:inputfisici}) for
$^{1}$H(p,$\nu e^+$)$^{2}$H 
and $^{14}$N(p,$\gamma$)$^{15}$O reaction rates, radiative opacity
$k_{\rm r}$, and microscopic diffusion velocities.  

\begin{figure}
\centering
\includegraphics[height=8.4cm,angle=-90]{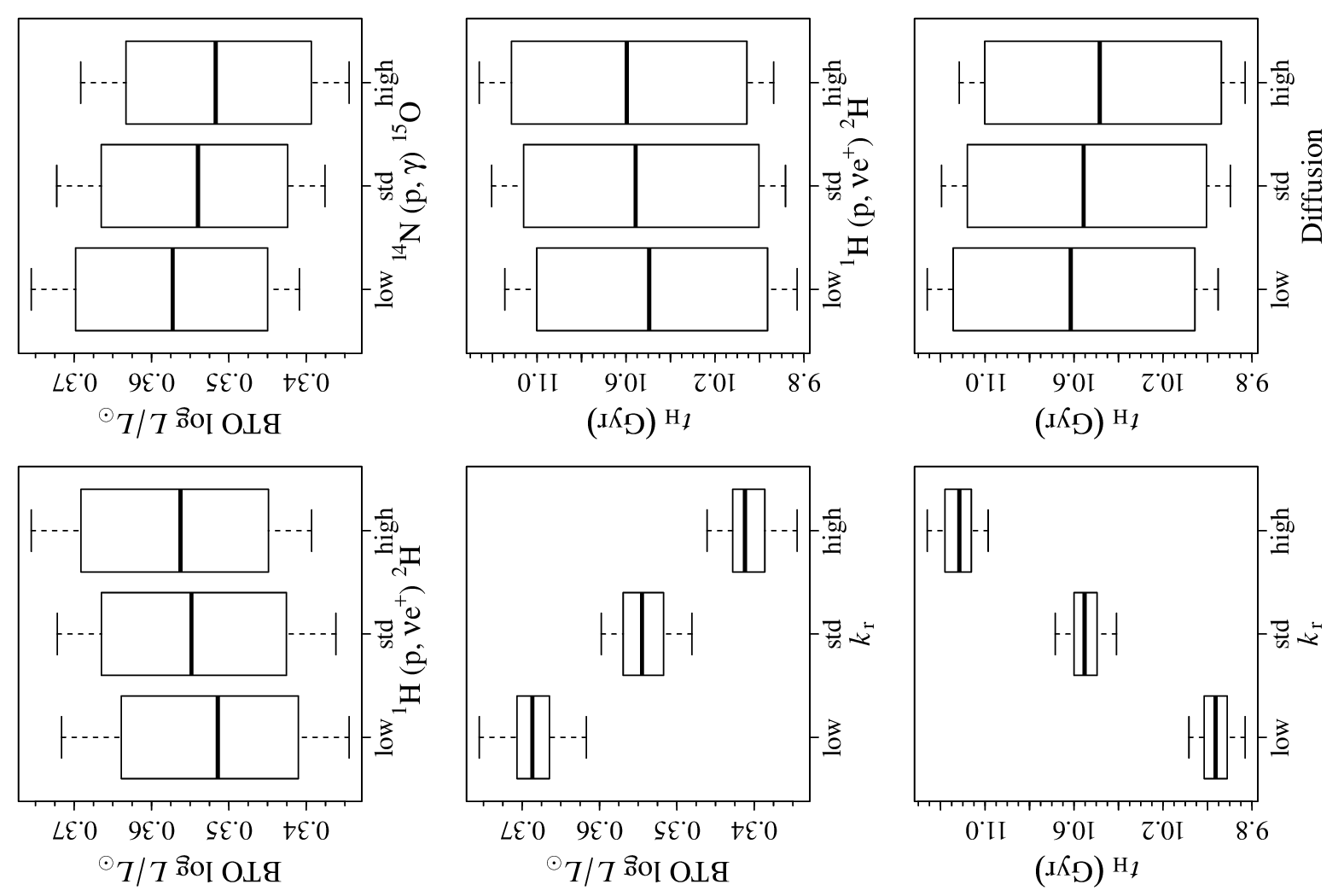}
\caption{Boxplot of the impact of the variation of the most important physical
inputs on $\log L_{\rm BTO}$ and on $t_{\rm H}$. The
black thick lines show the median of the data set, while the box  
marks the interquartile range, i.e. it extends form the 25th to the 
75th percentile of data. The  whiskers extend from the box until the extreme
data. For each parameter $p_i$, the labels low, std and high refer to the
values 1.00 - 
$\Delta p_i$, 1.00, and 1.00 + $\Delta p_i$.  }
\label{fig:box-plot-LTO-tHc}
\end{figure}

The regression model coefficients, along  with their statistical significance, 
are listed in Tables \ref{table:modLTO} and \ref{table:modtHc}. 
In the first two columns of the Tables we report the
least-squares estimates of the regression coefficients and their errors; in the
third column we report the $t$-statistic for the tests of the statistical
significance of the covariates. 
In the last row of the Tables we report the residual standard error of the fit
$\sigma$ and the value of the squared multiple correlation coefficient $R^2$,
i.e. the fraction of the variance of the data explained by the model, defined
as: 
\begin{equation}
R^2 = 1 - \frac{\sum (\hat y_i - y_i)^2}{\sum (y_i - \bar y)^2} 
\end{equation}
where $y_i$ are the values of the dependent variable, $\bar y$ their mean
value, and $\hat y_i$ the values predicted by the linear model.

All the 
tests reach a very high statistical significance ($p$-values $<2 \times
10^{-16}$).
The physical relevance of the different covariates can be assessed by looking
at Fig.~\ref{fig:box-plot-LTO-tHc} where we show the boxplots highlighting
the influence of the variation of the inputs among the three values
chosen for the calculations. 
For each parameter $p_i$ the boxplots are a convenient way to summarize the
variability of the data subsetted according to the three values of $p_i$
($1.00 - \Delta p_i$, 1.00, and $1.00 + \Delta p_i$, labeled as low, std, and
high in the plots).  
The black thick lines show the median of the data set, while the box 
marks the interquartile range, i.e. it extends form the 25th to the 
75th percentile of data. The  whiskers extend from the box until the extreme
data,  
the bottom whisker ranges from the sample minimum to the first quartile while
the  
top whisker from the third quartile to the sample maximum.  
While the position of the medians are related to the effect of the parameter 
in study in each plot, the extension of the box and whiskers are due to the
variation of all the other parameters. The larger the separation of the medians 
with respect to the dimension of the boxes and the greater the  importance 
of a given parameter.

The Figures show that the radiative opacity
uncertainty largely dominates for both $L_{\rm BTO}$ and $t_{\rm H}$. 
For $L_{\rm BTO}$ the impact of 
an increase of 5\% in $k_{\rm r}$ is $\Delta \log L_{\rm BTO}/L_{\sun} = 0.05 \times
(-0.276) = -0.0138$ dex, while the impact of the variation of the second most
important input -- 
i.e. $^{14}$N(p,$\gamma$)$^{15}$O reaction rate -- is only $\Delta
\log L_{\rm BTO}/L_{\sun} = -0.0028$ dex.
The sum of the effects in the statistical models does not account for the
whole variation of the data, due to the
presence of the random variation extracted in the residuals of the statistical
models. 

In the case of $t_{\rm H}$ the dominance of
$k_{\rm r}$ is even larger since its variation accounts for $\Delta
t_{\rm H} = 0.579$ Gyr, and the second most important input, the
microscopic diffusion velocities, for $\Delta
t_{\rm H} = 0.065$ Gyr.  

\begin{table}[ht]
\centering 
\caption{Fit of BTO log-luminosity (dex). }
\label{table:modLTO}
\centering
\begin{tabular}{lrrrr}
  \hline\hline
 & Estimate & Std. Error & $t$ value & Impact\\ 
&                  &            &           & (dex)\\ 
  \hline
  $\beta_0$        & $6.00 \times 10^{-1}$ & $3.43
  \times 10^{-3}$ & 174.91 & \\ 
 $\beta_1$ (pp) & $6.76 \times 10^{-2}$ &
  $2.81 \times 10^{-3}$ & 24.06 & 0.0020\\ 
  $\beta_2$ ($^{14}$N) &$-2.80 \times 10^{-2}$ &
  $8.43 \times 10^{-4}$ & -33.25 & -0.0028\\ 
 $\beta_3$  ($k_{\rm r}$) & $-2.76 \times 10^{-1}$ & $1.69 \times
  10^{-3}$ & -163.45  & -0.0138\\ 
 $\beta_4$ ($v_{\rm d}$)& $-9.56 \times 10^{-3}$ & $5.62 \times 10^{-4}$
 & -17.00 & -0.0014\\ 
   \hline
\multicolumn{5}{c}{$\sigma = 6.2 \times 10^{-4}$ dex; $R^2$ = 0.9974}\\
   \hline
\end{tabular}
\tablefoot{In the first two columns: least-squares
estimates of the regression coefficients and their errors;   
third column: $t$-statistic for the tests of the statistical 
significance of the covariates. 
In the last column is reported the physical impact of the
  variation $\Delta p_i$ of the various inputs. 
All the $p$-values of the tests are $<2 \times
10^{-16}$. The 
residual standard error $\sigma$ and the squared multiple correlation
coefficient $R^2$ are reported in the last row. }
\end{table}

\begin{table}[ht]
\centering
\caption{Fit of central hydrogen exhaustion time (Gyr). }
\label{table:modtHc}
\centering
\begin{tabular}{lrrrr}
  \hline\hline
 & Estimate & Std. Error & $t$ value & Impact\\ 
 &           &            &           & (Gyr)\\ 
  \hline
  $\beta_0$ &         -2.29 & 0.0421 & -54.28 & \\ 
    $\beta_1$ (pp) & 1.81   & 0.0345  & 52.46 & 0.054
  \\ 
  $\beta_2$ ($^{14}$N) &  -0.139 & 0.0104 & -13.39 & -0.014
  \\ 
   $\beta_3$ ($k_{\rm r}$) & 11.6   & 0.0207  & 559.81 & 0.579\\ 
  $\beta_4$ ($v_{\rm d}$)  &  -0.432 & 0.00690 & -62.56 & -0.065\\ 
 \hline
\multicolumn{5}{c}{$\sigma = 0.0076$ Gyr; $R^2$ = 0.9998}\\
   \hline
\end{tabular}
\tablefoot{All the $p$-values of
the tests are $<2 \times 10^{-16}$. The column
legend is the same as in Table~\ref{table:modLTO}.}
\end{table}

\begin{figure}
\centering
\includegraphics[height=8.4cm,angle=-90]{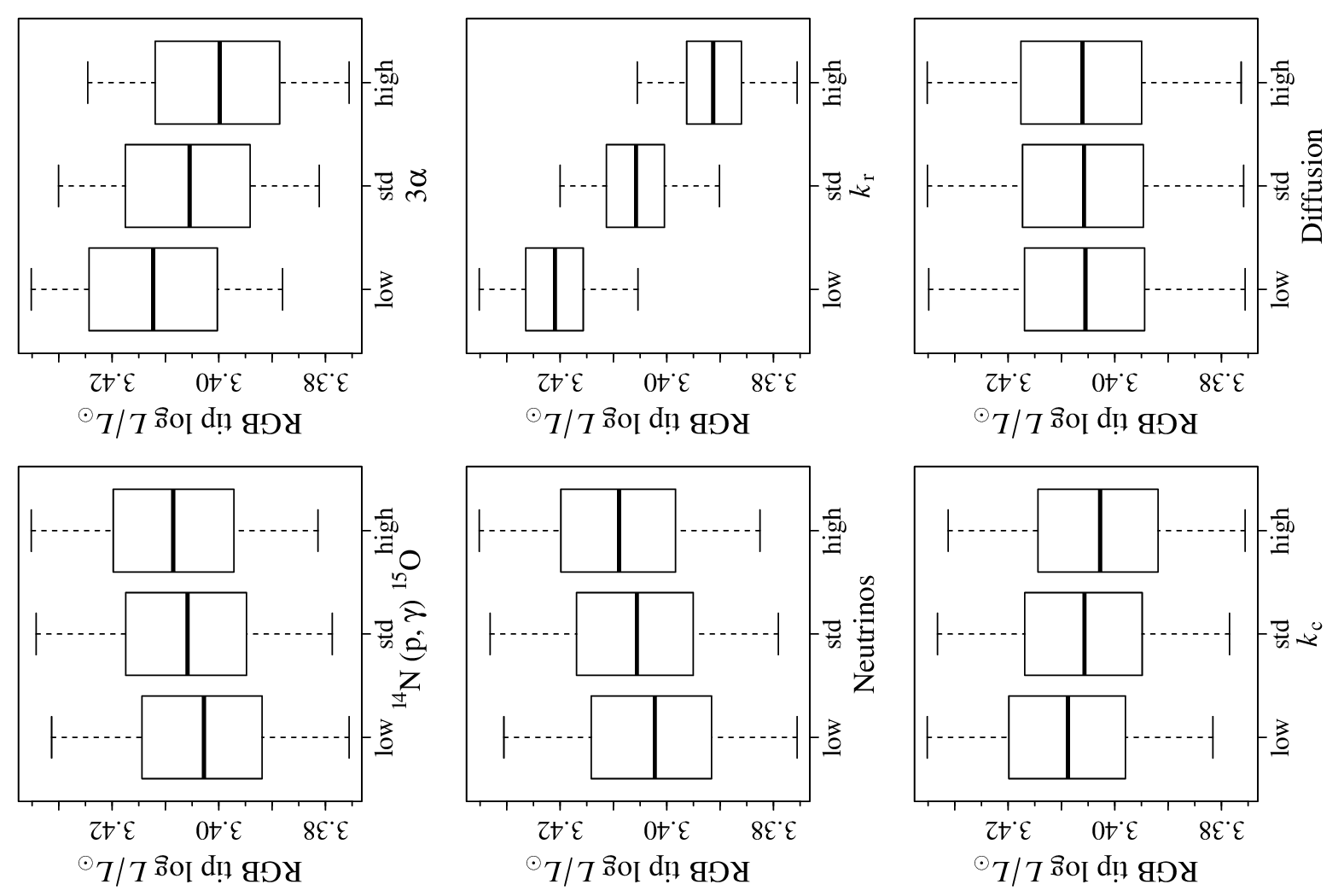}
\caption{Boxplot of the impact of the variation of the most important physical
inputs on $L_{\rm tip}$. }
\label{fig:box-plot-L-RGBt}
\end{figure}

\begin{figure}
\centering
\includegraphics[height=8.4cm,angle=-90]{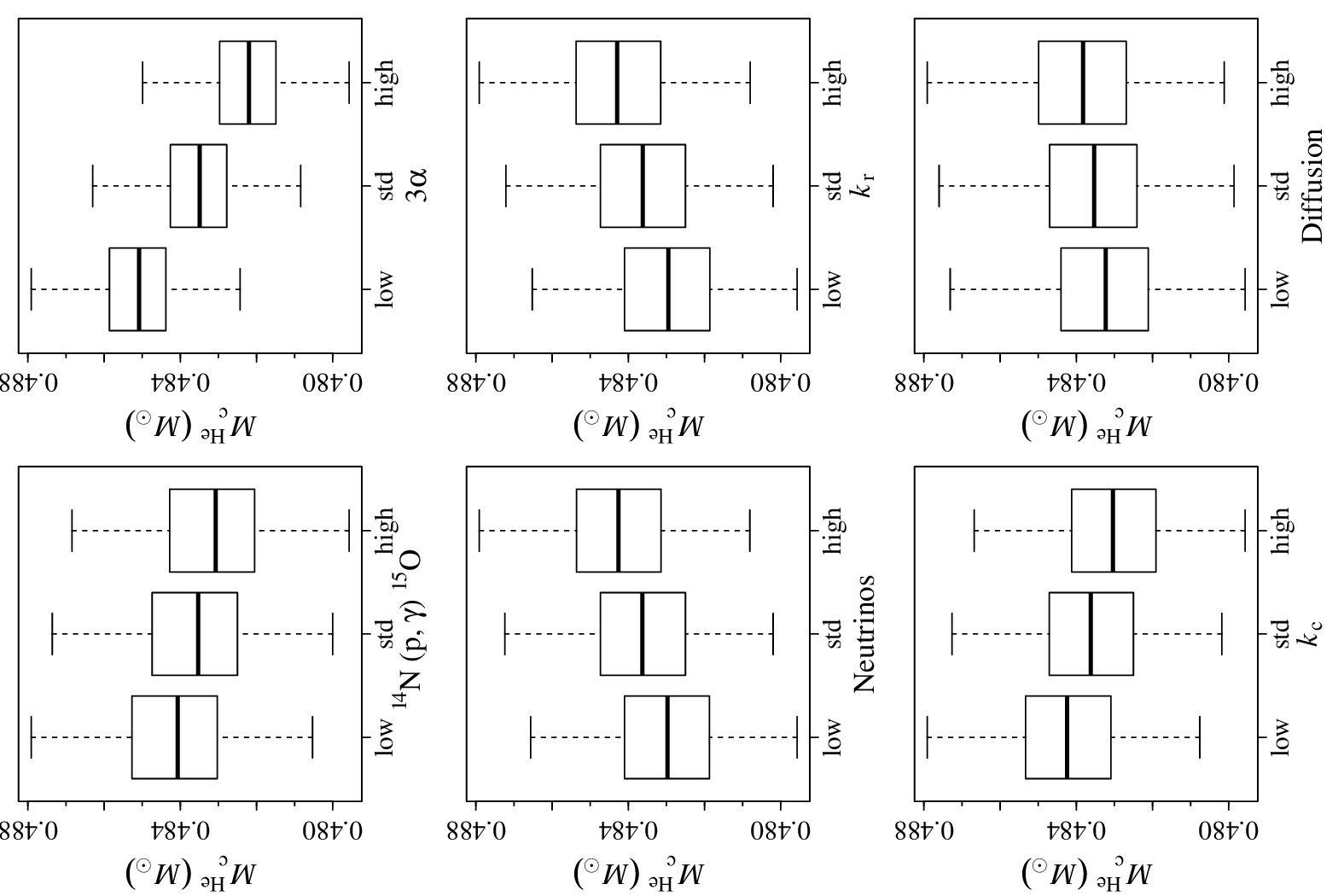}
\caption{Boxplot of the impact of the variation of the most important physical
inputs on $M_{\rm c}^{\rm He}$. }
\label{fig:box-plot-mche}
\end{figure}

\begin{figure}
\centering
\includegraphics[height=8.4cm,angle=-90]{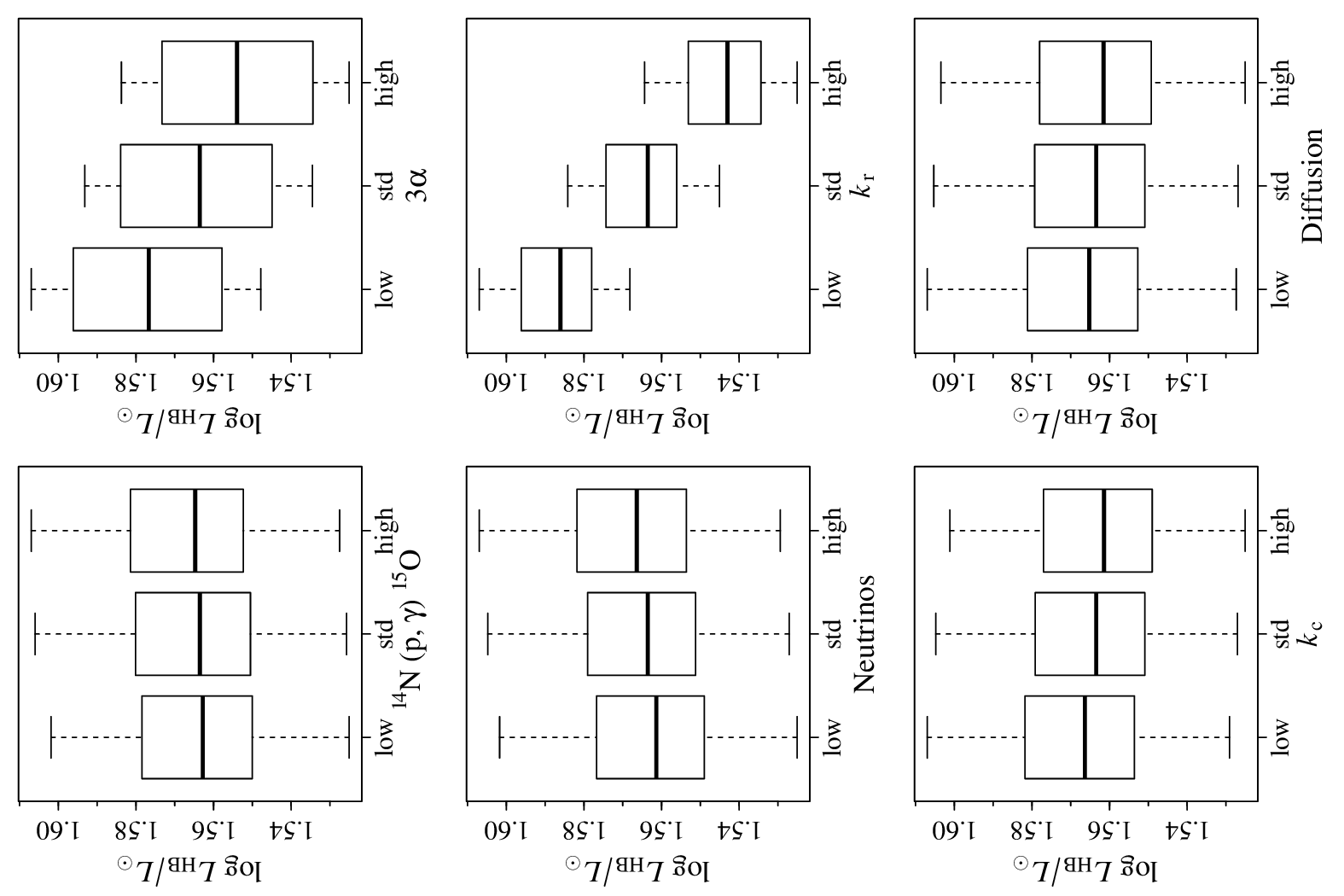}
\caption{Boxplot of the impact of the variation of the most important physical
inputs on $L_{\rm HB}$.  }
\label{fig:box-plot-L-383}
\end{figure}

In the cases of $\log L_{\rm tip}$, $\log L_{\rm HB}$ and $M_{\rm c}^{\rm He}$ 
the linear models were:
\begin{equation}
\log L_{\rm tip}, \log L_{\rm HB}, M_{\rm c}^{\rm He}  = \beta_0 + 
\sum_{i=1}^7 \beta_i \; p_i
\end{equation}
where the three additional physical
inputs considered are respectively the triple-$\alpha$ reaction rate, the
neutrino 
emission rate, and the conductive opacity 
$k_{\rm c}$ (see Table \ref{table:inputfisici}).
The regression model coefficients, along  with their
statistical significance, are listed in Tables \ref{table:modRGBt},
\ref{table:modmche}, and
\ref{table:modHB383}. 
The boxplots of the influence of the various inputs on the calculations are
presented in
Figs.~\ref{fig:box-plot-L-RGBt}, \ref{fig:box-plot-mche}, and
\ref{fig:box-plot-L-383}. 

For $\log L_{\rm tip}$ and $\log L_{\rm HB}$ the most important factor is
again the radiative opacity $k_{\rm r}$, although 
its impact is not as dominant as in the previous cases.  
For instance, in the case of $L_{\rm tip}$, 
the impact of an increase of 5\% in $k_{\rm r}$ is $\Delta \log L_{\rm
  tip}/L_{\sun} = -0.0149$ dex, 
while the variation of the second most important input,
the triple-$\alpha$ reaction rate, accounts for $\Delta
\log L_{\rm tip}/L_{\sun} = -0.0061$ dex, i.e. about 40\% of the $k_{\rm r}$
effect. 

For $L_{\rm HB}$ the effect of the variation of $k_{\rm r}$ is $\Delta \log
L_{\rm HB}/L_{\sun} = -0.0216$ 
dex, while the one of triple-$\alpha$ is $\Delta \log L_{\rm HB}/L_{\sun} = -0.0115$
dex, i.e. about 55\% of the previous one. 
  
Regarding the helium core mass, the main variation is due to
triple-$\alpha$ reaction rate: an increase of 20\% of this value accounts
 for $\Delta M_{\rm c}^{\rm He} = -0.00144$ $M_{\sun}$; the effect due to the
 uncertainty on 
 neutrino emission rate, radiative and conductive opacities are all of the
 same order, i.e.  
  about 45\% of the triple-$\alpha$ one each, while 
 $^{14}$N(p,$\gamma$)$^{15}$O reaction rate uncertainty accounts for about 35\% of
 the triple-$\alpha$ effect. 

\begin{table}[ht]
\centering
\caption{Fit of RGB tip log-luminosity (dex). }
\label{table:modRGBt}
\centering
\begin{tabular}{lrrrr}
  \hline\hline
 & Estimate & Std. Error & $t$ value &Impact\\ 
        &           &            &           & (dex)\\ 
  \hline
  $\beta_0$ & 3.68 & $6.41 \times 10^{-4}$  & 5743.68 &\\ 
  $\beta_1$ (pp) & $-2.56\times 10^{-3}$ & $4.11 \times 10^{-4}$ &
  -6.23\tablefootmark{*} & 0.0000\\  
  $\beta_2$ ($^{14}$N) & $2.74 \times 10^{-2}$ &
  $1.23 \times 10^{-4}$ & 222.28 & 0.0027\\ 
  $\beta_3$ ($k_{\rm r}$)      & $-2.98 \times 10^{-1}$ & $2.47 \times
  10^{-4}$ & -1209.61 & -0.0149\\ 
  $\beta_4$ ($v_{\rm d}$)      & $1.72 \times 10^{-3}$ & $8.22 \times 10^{-5}$
  & 20.91 & 0.0002\\ 
  $\beta_5$ (3$\alpha$)       & $-3.05 \times 10^{-2}$ & $6.16 \times
  10^{-5}$ & -494.01 & -0.0061\\ 
  $\beta_6$ ($\nu$)        & $8.22 \times 10^{-2}$ & $3.08 \times 10^{-4}$
  & 266.77 & 0.0033\\ 
  $\beta_7$ ($k_{\rm c}$) &     $-5.81\times 10^{-2}$ & $2.47 \times
  10^{-4}$ & -235.71 & -0.0029\\ 
   \hline
\multicolumn{5}{c}{$\sigma = 4.7 \times 10^{-4}$ dex; $R^2$ = 0.9988}\\
   \hline
\end{tabular}
\tablefoot{The $p$-values of the
 tests are $<2 \times 10^{-16}$, if not differently specified. The column
legend is the same as in Table~\ref{table:modLTO}. \\
\tablefoottext{*}{$p$-value $= 5.6 \times 10^{-10}$.}\\
}
\end{table}

\begin{table}[ht]
\begin{center}
\caption{Fit of helium core mass $M_{\rm c}^{\rm He}$ ($M_{\sun}$). }
\label{table:modmche}
\begin{tabular}{lrrrrr}
  \hline\hline
& Estimate & Std. Error & $t$ value & Impact\\
       &           &            &           & ($M_{\sun}$)\\
  \hline
  $\beta_0$ & $4.78 \times 10^{-1}$  &  $1.03 \times 10^{-4}$ & 4635.03 &\\
  $\beta_1$ (pp) & $-1.20 \times 10^{-3}$ & $6.61
  \times 10^{-5}$ & -18.16 &   0.00004\\
  $\beta_2$ ($^{14}$N) & $-4.95 \times 10^{-3}$ & $1.98
  \times 10^{-5}$ & -249.69 & -0.00049\\
  $\beta_3$ ($k_{\rm r}$) & $1.31 \times 10^{-2}$  & $3.96 \times 10^{-5}$ &
  329.33 & 0.00065\\
  $\beta_4$ ($v_{\rm d}$) & $1.93 \times 10^{-3}$  & $1.32 \times 10^{-5}$ &
  146.35 & 0.00029\\
  $\beta_5$ (3$\alpha$) & $-7.20\times 10^{-3}$  & $9.91 \times 10^{-6}$
  & -726.95 & -0.00144\\
  $\beta_6$ ($\nu$) & $1.61 \times 10^{-2}$  & $4.96 \times 10^{-5}$ &
  324.28 & 0.00064\\
  $\beta_7$ ($k_{\rm c}$) & $-1.20 \times 10^{-2}$ & $3.96 \times 10^{-5}$ &
  -303.07 & -0.00060\\
   \hline
\multicolumn{5}{c}{$\sigma = 7.6 \times 10^{-5}$ $M_{\sun}$; $R^2$ = 0.9976}\\
   \hline
\end{tabular}
\tablefoot{All the
  $p$-values of 
the tests are $<2 \times 10^{-16}$. The column
legend is the same as in Table~\ref{table:modLTO}.}
\end{center}
\end{table}

\begin{table}[ht]
\centering
\caption{Fit of the ZAHB log-luminosity $\log L_{\rm HB}$ (dex) at $\log T_{\rm
    eff}$ = 3.83. }
\label{table:modHB383}
\centering
\begin{tabular}{lrrrr}
  \hline\hline
 & Estimate & Std. Error & $t$ value& Impact\\ 
        &           &            &           & (dex)\\
  \hline
  $\beta_0$    &  2.03                  & $1.20 \times 10^{-3}$ & 1696.88 & \\ 
  $\beta_1$ (pp)       & $6.93 \times 10^{-3}$  &
  $7.68 \times 10^{-4}$ & 9.03 & 0.0002\\ 
  $\beta_2$ ($^{14}$N)       & $1.20 \times 10^{-2}$  &
  $2.30 \times 10^{-4}$ & 52.04 & 0.0012\\ 
  $\beta_3$ ($k_{\rm r}$)       & $-4.33 \times 10^{-1}$ & $4.61 \times
  10^{-4}$ & -939.64 & -0.0216\\ 
  $\beta_4$  ($v_{\rm d}$)       & $-1.12 \times 10^{-2}$ & $1.54 \times
  10^{-4}$ & -73.17 & -0.0017\\ 
  $\beta_5$ (3$\alpha$)       & $-5.78 \times 10^{-2}$ & $1.15 \times
  10^{-4}$ & -501.83 & -0.0115\\ 
  $\beta_6$ ($\nu$)       & $5.77 \times 10^{-2}$  & $5.76 \times 10^{-4}$
  & 100.22 & 0.0023\\ 
  $\beta_7$ ($k_{\rm c}$)       & $-4.36 \times 10^{-2}$ & $4.61 \times
  10^{-4}$ & -94.70 & -0.0022\\ 
   \hline
\multicolumn{5}{c}{$\sigma = 8.8 \times 10^{-4}$ dex; $R^2$ = 0.9981}\\
  \hline
\end{tabular}
\tablefoot{All the 
  $p$-values of the 
  tests are $<2 \times 10^{-16}$.  The column
legend is the same as in Table~\ref{table:modLTO}.}
\end{table}

The statistical models presented in this Section rely on the  
assumption of the linearity of the output of the calculations
with respect to the perturbations of the physical inputs. 
As a check of this
hypothesis, we made some additional runs varying only one parameter at a time,
in a 
set of 9 points spanning the assumed range of uncertainty. The procedure is
fully justified by the lack of interaction among physical inputs in the linear
models 
presented in Tables \ref{table:modLTO} - \ref{table:modHB383}. In fact in
presence of a 
relevant interaction, the values of the multiple correlation coefficient
$R^2$  of the fits would be 
much smaller than the ones found in the models, which had a minimum
value of 0.997 for $L_{\rm BTO}$, i.e. the 99.7\% of the variance of the data
are explained by the model.
 
\begin{figure*}
\centering
\includegraphics[width=10cm,angle=-90]{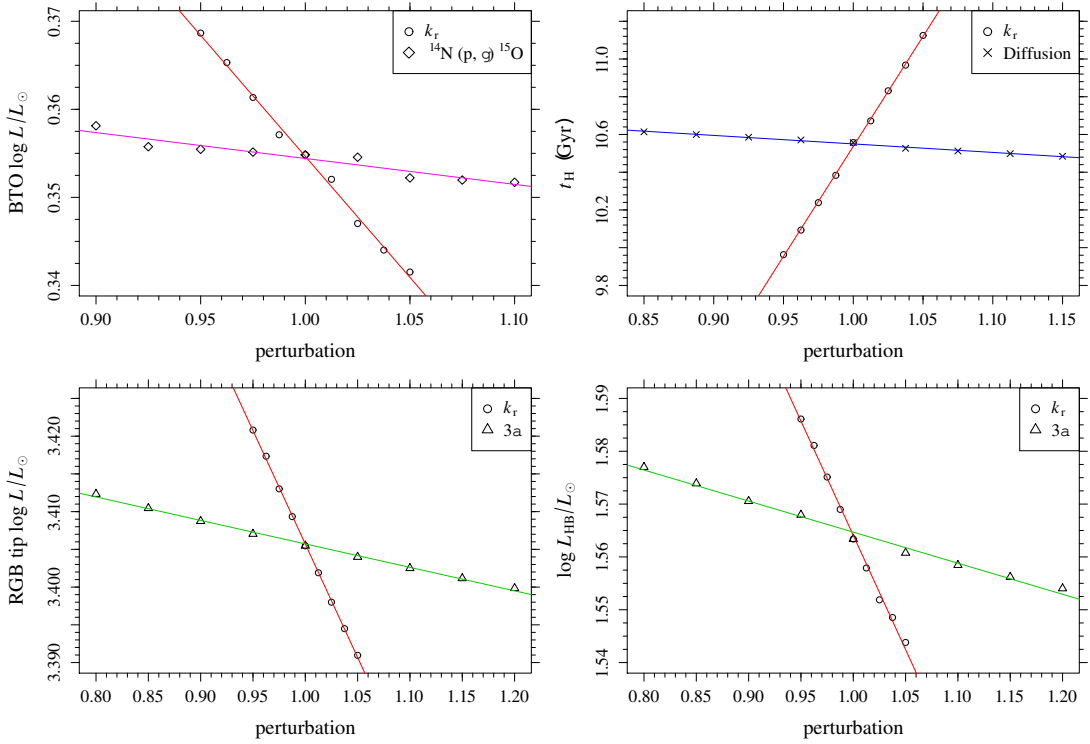}
\caption{Study of the linearity of the outputs of the calculations with respect
to the variation of the physical inputs in their range of uncertainty. The
symbols are the results of the calculations, while the lines are the linear fit
of each set of values. Only the two most important inputs are shown in each
panel. }
\label{fig:linearity}
\end{figure*}

In Fig.~\ref{fig:linearity} we present a graphic test of linearity. 
The four panels show, for each evolutionary feature studied in the
Section, the results of the calculations for the variation of the two most
important physical inputs with superimposed the linear fit of the different
trends. In all 
the cases the assumption of linearity is respected with high accuracy. 

Since the hypotheses of linearity in the inputs and their
independence 
hold, the statistical models presented in this Section can be safely used to
interpolate 
the effect of a perturbation of the physical inputs in the assumed range.
   
In the case of the central hydrogen exhaustion (i.e. X$_{\rm c}$ = 0), 
the effect of the radiative opacity variation will equate the one due
to the second  
most important contribution, namely that due to the microscopic velocity
variation,   
for a perturbation estimated as the product of assumed perturbation by the
ratio of the effect of the two inputs as resulted by the model, i.e. $0.05
\times 0.065/0.58 = 0.0056$. This means that in order to reduce the
uncertainty in the predicted  
main sequence lifetime due to the radiative opacity at the level of the second
largest contribution, the  
uncertainty affecting the Rosseland mean opacity should decrease from the
currently  
adopted 5\% down to 0.56\%. 
For the turn-off $L_{\rm BTO}$ and ZAHB $L_{\rm HB}$ luminosities, the
radiative opacity  
remains the most important uncertainty source for assumed perturbation greater
or equal to 
 $0.05 \times 0.0028/0.014 = 0.01$ and $0.05 \times 0.012/0.022 = 0.027$,
respectively.  
Thus, in order to reduce the uncertainty in the turn-off luminosity caused by
the radiative opacity to  
the same order of that caused by the second most important contribution,
i.e. the  
$^{14}$N(p,$\gamma$)$^{15}$O reaction rate, the uncertainty of the former
should be decreased  
from 5\% to 1\%. Finally, the radiative opacity should be known with a
precision  
better than about 2.7\% to lead a variation in the predicted ZAHB luminosity
lower than the second  
largest contribution, i.e. the triple-$ \alpha $ reaction rate.

\section{Global physical uncertainty in stellar isochrones}
\label{sec:iso}

The direct theoretical counterparts of the observed color-magnitude diagrams of 
simple stellar populations are
isochrones rather than tracks with fixed mass.  
For this reason, we extended the previous statistical analysis of the physical 
uncertainties affecting stellar models to isochrones with age in the range
8-14 Gyr, suitable  
for galactic globular clusters, with time steps of 1.0 Gyr. 

Since we were interested mainly in studying the variation near the turn-off
region,  
we performed calculations varying only the four physical inputs $p_1, \ldots,
p_4$ that can 
influence the evolution until this phase, as shown in the previous section.
Thus, for each set of perturbed input physics, i.e. for each set of multipliers 
values $p_1, \ldots, p_4$, we computed a grid of stellar tracks with
different masses,  
for a total of $ 3^{4}=81 $ grids. Each grid contains 12 stellar models with
mass values chosen to 
 accurately reconstruct the zone near the BTO, namely
  $M = 0.40, 0.50, 0.60, 0.70, 0.80, 0.85, 0.87, 0.90, 0.92, 0.95, 1.00, 1.10$
 $M_{\sun}$.  
Then we computed 972 stellar tracks and 567 isochrones with fixed chemical
composition  
($Z = 0.006$, $Y = 0.26$) and mixing-length parameter ($\alpha_{\rm ml}= 1.90$).

\begin{figure}
\centering
\includegraphics[width=6.2cm,angle=-90]{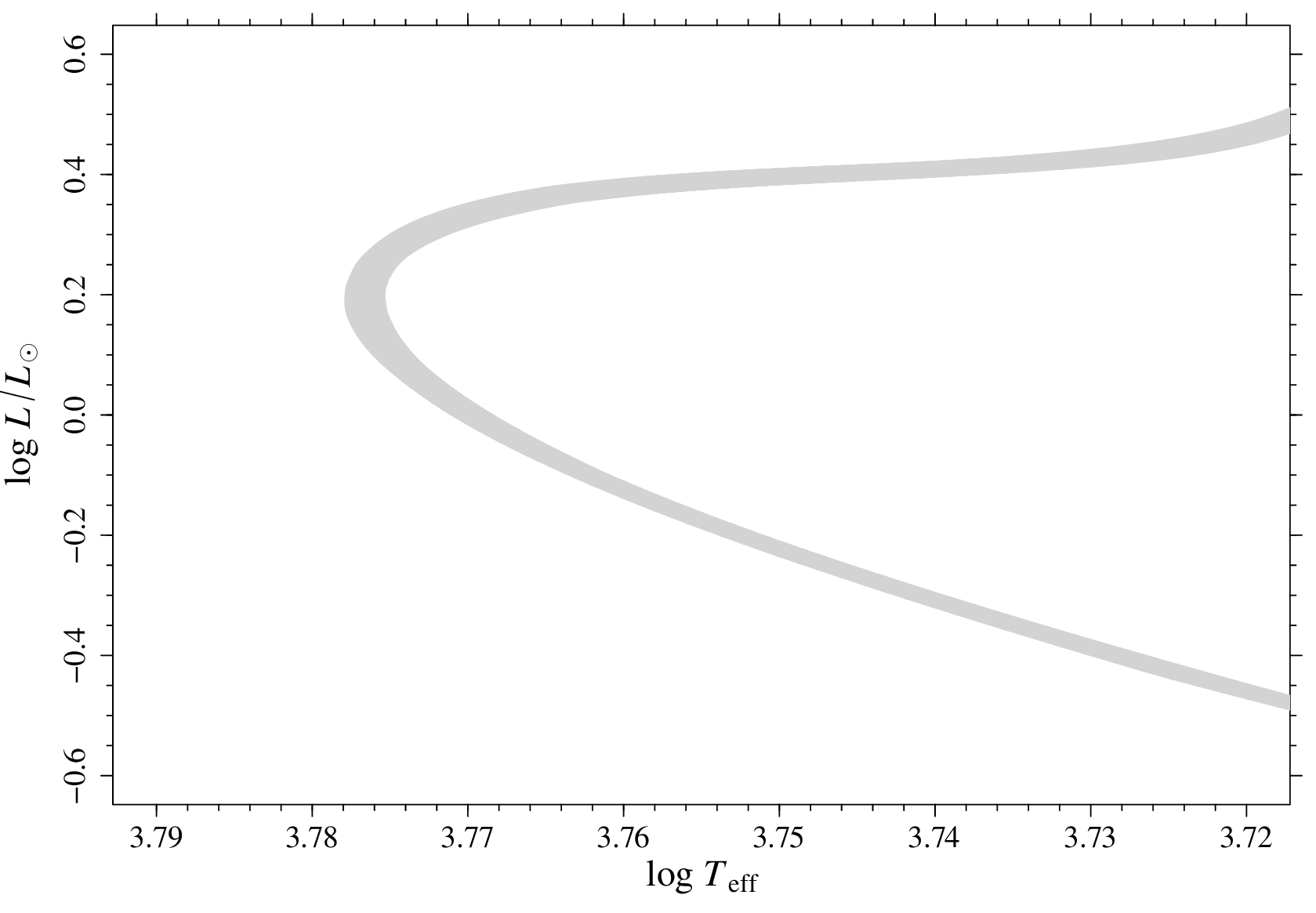}
\caption{HR diagram showing the error stripe due to the variation of the
  physical inputs for 
a 12.0 Gyr isochrone (zoom of the TO region).} 
\label{fig:nastro-iso}
\end{figure}

The global effect of the combined variation of the four selected inputs on the
12.0 Gyr isochrone in the ($\log T_{\rm eff}$, $\log L/L_{\sun}$) plane is
displayed in Fig.~\ref{fig:nastro-iso}. The stripe around the  
isochrone computed with standard inputs represents the region of the plane
spanned by the perturbed stellar models.

\begin{figure*}
\centering
\includegraphics[height=14cm,angle=-90]{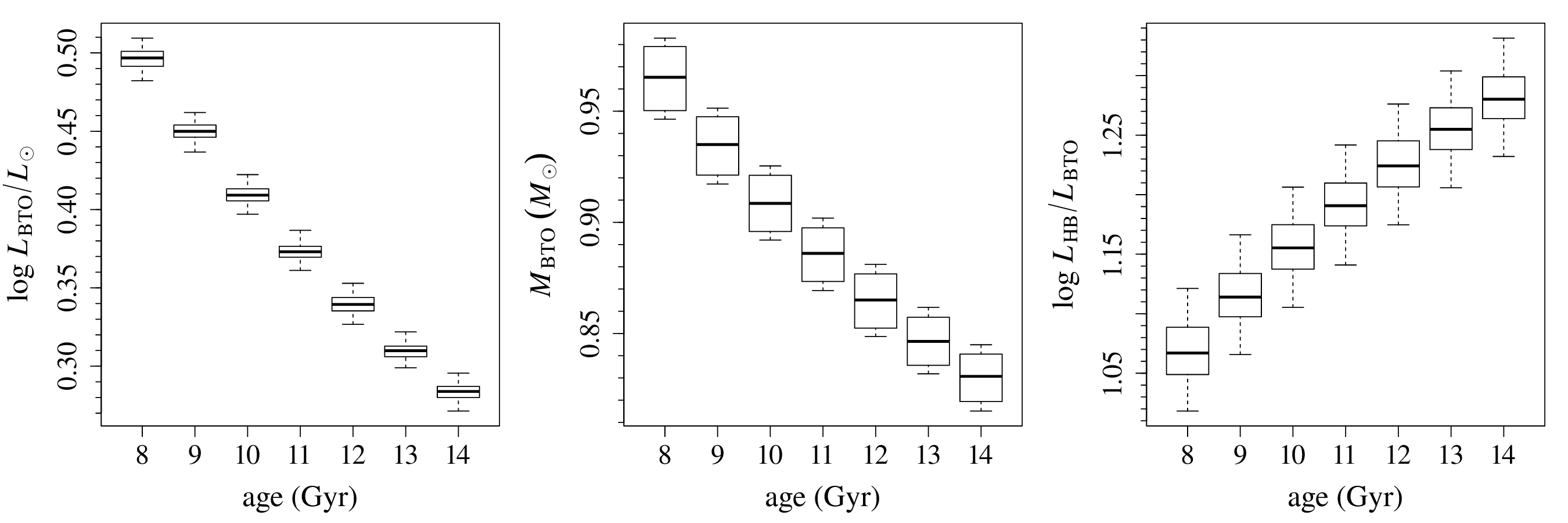}
\caption{Left panel: boxplots for BTO luminosity due to the variation of the
selected physical inputs as a function of the isochrone age in Gyr. Central
panel: same plot for mass at BTO. Right panel: same plot for difference of HB
and BTO log luminosity. }
\label{fig:iso-LM}
\end{figure*}

Figure \ref{fig:iso-LM}  shows the boxplots of
  three selected  quantities as a function of 
 the isochrone age, namely the turn-off luminosity ($L_{\rm BTO}^{\rm iso}$) 
 and mass ($M_{\rm BTO}^{\rm  iso}$), and the differences between
$\log L_{\rm HB}$ and $\log L_{\rm BTO}^{\rm iso}$. The last one
is the theoretical  
counterpart of the $ \Delta V(TO-HB) $, i.e. the visual magnitude difference
between  
turn-off and horizontal branch regions, used as age indicator in the vertical
method  
technique.  
 
Table \ref{table:incertezzeisocrona} lists the half-width of the variation
range in
predictions of the selected quantities for our reference stellar isochrone of
12 Gyr 
 due to current input physics uncertainties.    
The turn-off log luminosity $\log L_{\rm BTO}^{\rm iso}$ and mass
 $M_{\rm BTO}^{\rm  iso}$ vary of $\pm 0.013$ dex 
 and $\pm 0.015$ $M_{\sun}$, respectively, while
the $\log L_{HB}/L_{\rm BTO}^{\rm iso}$ vary of $\pm 0.05$ dex.  

\begin{table}[ht]
\centering
\caption{Range half-width of variation in theoretical predictions of selected
  quantities for our reference isochrone of 12 Gyr,
   with $Z = 0.006$ and $Y = 0.26$, due to input
physics uncertainties.}  
\label{table:incertezzeisocrona}
\centering
\begin{tabular}{lc}
  \hline\hline
 quantity & range half-width \\
 \hline
  $\log L^{iso}_{\rm BTO}$  &  0.013 dex \\
  $ M^{iso}_{\rm BTO}$  &  0.015 $M_{\sun}$ \\
  $\log L_{HB}/L^{iso}_{\rm BTO}$  &  0.05 dex \\
   \hline
\end{tabular}
\end{table}

\section{Statistical analysis of physical uncertainty in stellar isochrones}
\label{sec:statistic_isoc}

To explore the influence of the physical inputs variation on the reconstructed 
turn-off log-luminosity $\log L_{\rm BTO}^{\rm iso}$ and mass $M_{\rm BTO}^{\rm
  iso}$,  we chose to pool the results for different
ages removing the trend due to the age, increasing the possibility to detect
the effect of the perturbations due to a larger statistic.
The removal of the age trend was done by adapting a linear
model to either $\log L_{\rm BTO}^{\rm iso}$ and $M_{\rm BTO}^{\rm iso}$ using as
covariate a categorical variable, i.e. a variable which assumes different
values for each age. The variable is then used as in a classical 
ANOVA (i.e. analysis of variance) model to
filter out the variation due to the different isochrones ages. 
The residuals of the models were then regressed with respect to $p_1, \ldots,
p_4$.  
The procedure guarantees the removal of the mean trend from the data, i.e. every
set of residuals for each age has mean value of zero. 
The technique assumes that the effect of $p_i$ are the same at different ages
in  
the studied time interval, i.e. 8-14 Gyr.
 As a rapid check of the hypothesis, we performed two Bartlett tests of
homogeneity of variances of residuals among ages \citep{snedecor1989}. The
tests did not suggest any 
problem ($\log L_{\rm BTO}^{\rm  iso}$ $p$-value = 0.16,  $M_{\rm BTO}^{\rm iso}$
$p$-value = 0.26).

\begin{figure}
\centering \includegraphics[width=8.4cm,angle=-90]{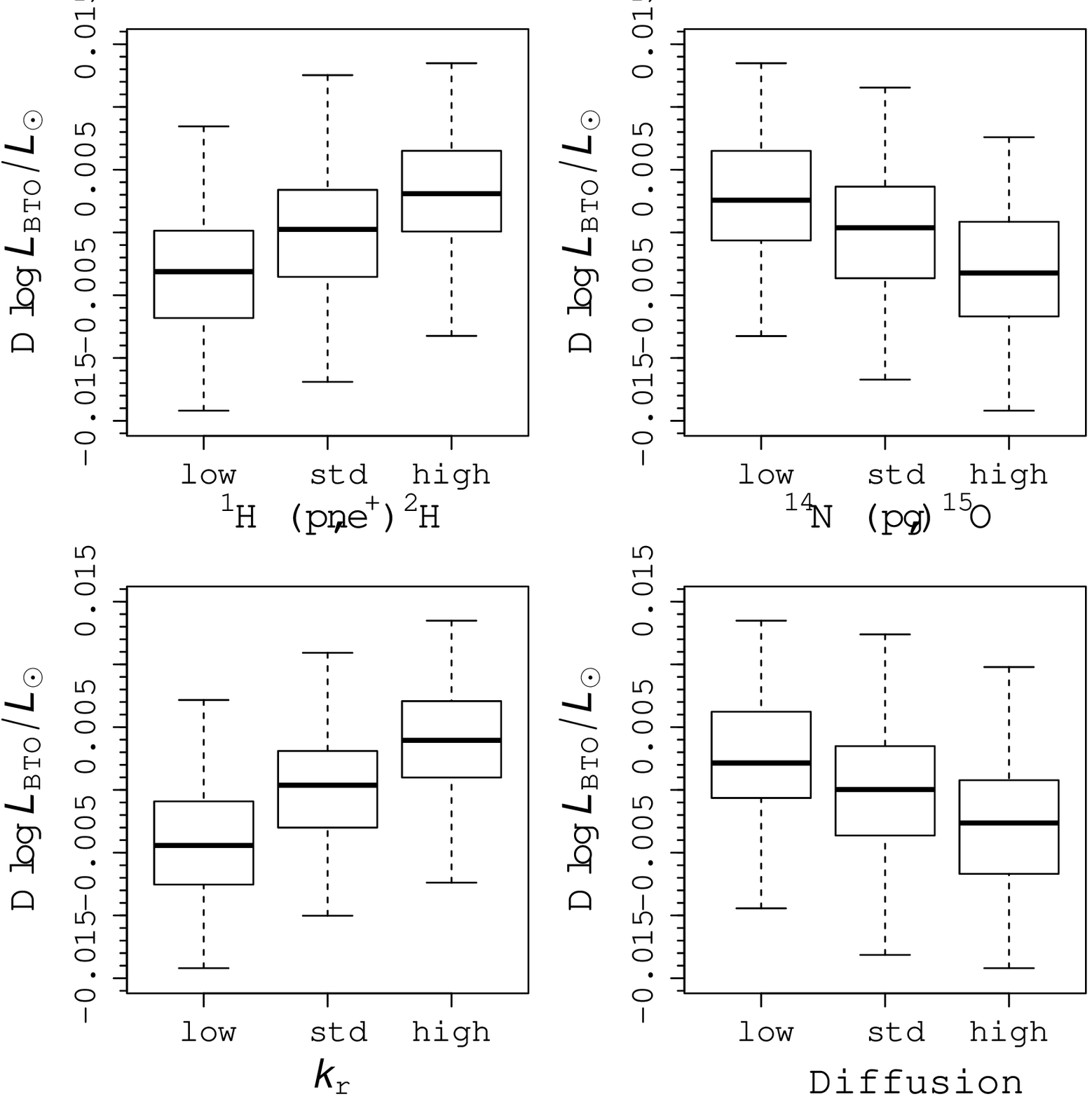}
\caption{Boxplot of the impact of the variation of the chosen physical inputs
  on the isochrone turn-off luminosity $L_{\rm BTO}^{\rm iso}$. The results for
  ages in the range [8 - 14] Gyr are pooled together, see text for details.}
\label{fig:iso-logL}
\end{figure}

\begin{figure}
\centering \includegraphics[width=8.4cm,angle=-90]{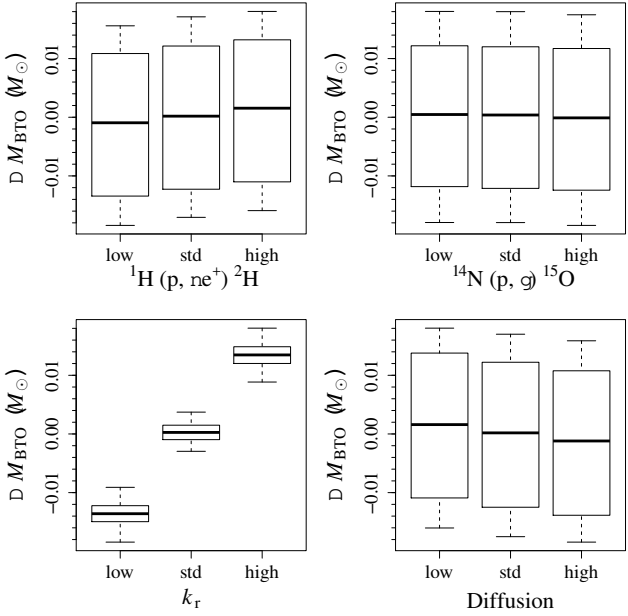}
\caption{Boxplot of the impact of the variation of the chosen physical inputs
  on the isochrone turn-off mass $M_{\rm BTO}^{\rm iso}$. The results for
  ages in the range [8 - 14] Gyr are pooled together, see text for details.}
\label{fig:iso-M}
\end{figure}

The results of the linear models for $\log L_{\rm BTO}^{\rm  iso}$ and $M_{\rm
  BTO}^{\rm iso}$ are in Table \ref{table:isoTOl} and
\ref{table:isoTOm}, respectively. In Figs.~\ref{fig:iso-logL} and
\ref{fig:iso-M} we present the boxplots which evidence
the influence of the variation of the physical inputs among the three values
chosen for the calculations. For the turn-off mass $M_{\rm BTO}^{\rm
  iso}$, the effect of 
the variation of the radiative opacity is largely dominant. 
An increase of 5\% in $k_{\rm r}$ produces a variation of $\Delta
M = 0.05 \times 
0.270 = 0.0135$ $M_{\sun}$, while the impact of the variation of the second most
important input -- 
i.e. microscopic diffusion velocities -- is only $\Delta M = -0.0014$
$M_{\sun}$. The total range of variation of $\Delta M_{\rm BTO}^{\rm iso}$ is
[-0.0184 - 0.0180] $M_{\sun}$.

For the luminosity of the turn-off $L_{\rm BTO}^{\rm iso}$, no physical input
definitely  
dominates on the others. The most important factor turned out to be radiative
opacity 
with a variation $\Delta \log L/L_{\sun} = 0.0042$ dex, followed by
$^{1}$H(p,$\nu e^+$)$^{2}$H 
reaction rate with a variation $\Delta \log L/L_{\sun} = 0.0031$ dex, 
$^{14}$N(p,$\gamma$)$^{15}$O reaction rate with $\Delta \log L/L_{\sun} =
-0.0029$ dex, and microscopic diffusion velocities with $\Delta \log L/L_{\sun} =
-0.0026$ dex. The total range of variation of $\Delta \log L_{\rm BTO}^{\rm iso}$ is
[-0.0142 - 0.0135] dex, about 2/3 of the one found in Sec.~\ref{sec:track}
for the turn-off luminosity $\log L_{\rm BTO}$ of the reference track.
 This lower uncertainty in the isochrones with respect to the
tracks used to build them is due to the fact that isochrones of the same age
but  
computed with different sets of multipliers $p_1, \ldots, p_4$ values have 
turn-off regions populated by different masses $M_{\rm BTO}^{\rm iso}$. 
As already shown, 
the variation of $k_{\rm r}$ splits the mass at BTO in three
separate sets for each value of the parameter (see Fig.~\ref{fig:iso-M}). 
The higher the value of $k_{\rm r}$, the more massive are the stellar models
at BTO. 
The luminosity of these models, which
are intrinsically 
higher, are depleted by the high value of the radiative opacity, as in the
third panel of first row in Fig.~\ref{fig:box-plot-LTO-tHc}. The opposite
behavior was evidenced for low value of $k_{\rm r}$. The net effect is a
shrinkage of the luminosity range spanned by the models.

The most important and common application of theoretical isochrones is 
the age estimate of stellar clusters. As
such, it would be of primary 
importance  
to evaluate the current uncertainty affecting the age inferred from the
comparison between  
 observed and predicted turn-off luminosity, in addition to the uncertainty in
 the  
 BTO luminosity at a given age analyzed above.
 The computed isochrones can be used for such purpose, even if a special care
is needed. In fact a direct modeling of the age with respect to 
$\log L_{\rm  BTO}^{\rm iso}$ is not feasible, since the age is used in the
calculation 
as a parameter with null uncertainty and it cannot be viewed as random
variable, as 
required for a linear model dependent variable.
Therefore we proceed in the following way. We found that
the relation between $\log L_{\rm BTO}^{\rm iso}$ and the age is well
described by the following linear model:
\begin{equation}
\log L_{\rm BTO}^{\rm iso} = \beta_0 + \beta_1 \; t_9 
\end{equation}
where $t_9 = \log t$, with $t$ the age of the isochrone in Gyr. The
parameters of the model are: $\beta_0 = 1.2871$, $\beta_1 = -0.8771$, with
$\sigma = 0.0056$ dex.
Since we do include explicitly the contribution of the parameter $p_1, \ldots,
p_4$, they contribute to the inflation of the residual standard error of
the model, reflecting our global uncertainty.
This model can be used to perform a reverse inference on
the value of the age, given $L_{\rm BTO}^{\rm iso}$. As described in detail in
Appendix \ref{app:intconf}, in the explored range [8 - 14] Gyr, a typical
uncertainty in the age is about $\pm$ 375 Myr.

With an analysis like the one described in the previous paragraph for  $L_{\rm
  BTO}^{\rm iso}$, we can
obtain a reverse inference of the uncertainty on age given the value of 
$\log L_{\rm HB}/L_{\rm BTO}^{\rm iso}$. 
The linear model adapted to data is:
\begin{equation}
\log L_{\rm HB}/L_{\rm BTO}^{\rm iso} = \beta_0 + \beta_1 \; t_9 
\end{equation}
where $t_9 = \log t$, with $t$ the age of the isochrone in Gyr. The
parameters of the model are: $\beta_0 = 0.2779$, $\beta_1 = 0.8771$, with
$\sigma = 0.024$ dex.
As a result of the large value of $\sigma$, a typical uncertainty in
age, given $\log L_{\rm  HB}/L_{\rm BTO}^{\rm iso}$, is of the order of $\pm$
1.25 Gyr.

\begin{table}[ht]
\centering
\caption{Fit of the isochrones BTO log-luminosity (dex). } 
\label{table:isoTOl}
\centering
\begin{tabular}{lrrrr}
  \hline\hline
 & Estimate & Std. Error & $t$ value &Impact\\
    &  &    &   & (dex)\\ 
\hline 
$\beta_0$ & $-1.40
  \times 10^{-1}$ & $2.72 \times 10^{-3}$ & -51.51 &\\ 
$\beta_1$ (pp)& $1.04 \times
  10^{-1}$ & $2.23 \times 10^{-3}$ & 46.60 & 0.0031\\ 
$\beta_2$ ($^{14}$N) & $-2.95 \times
  10^{-2}$ & $6.68 \times 10^{-4}$ & -44.13 & -0.0029\\ 
$\beta_3$ ($k_{\rm r}$) & $8.34 \times
  10^{-2}$ & $1.34 \times 10^{-3}$ & 62.43 & 0.0042\\ 
$\beta_4$ ($v_{\rm d}$) & $-1.77 \times
  10^{-2}$ & $4.45 \times 10^{-4}$ & -39.65 & -0.0026\\ 
  \hline
\multicolumn{5}{c}{$\sigma = 0.0013$ dex; $R^2$ = 0.945}\\
  \hline
\end{tabular}
\tablefoot{All the $p$-values of the
tests  are $<2 \times 10^{-16}$. The column legend is the same as
in Table~\ref{table:modLTO}.}
\end{table}

\begin{table}[ht]
\centering
\caption{Fit of mass at BTO ($M_{\sun}$). }
\label{table:isoTOm}
\centering
\begin{tabular}{lrrrr}
  \hline \hline
& Estimate & Std. Error & $t$ value & Impact\\
 & & & & ($M_{\sun}$)\\ 
\hline $\beta_0$ & $-3.00
  \times 10^{-1}$ & $2.08 \times 10^{-3}$ & -143.97 &\\ 
$\beta_1$ (pp)& $4.13 \times 10^{-2}$ & $1.71 \times 10^{-3}$ & 24.23 & 0.0012\\ 
$\beta_2$ ($^{14}$N) & $-2.40 \times 10^{-4}$ & $5.12 \times 10^{-4}$ &
  -4.69\tablefootmark{*} & -0.0002\\ 
$\beta_3$ ($k_{\rm r}$) & $2.70 \times 10^{-1}$ & $1.02 \times 10^{-3}$ & 264.03 & 0.0135\\ 
$\beta_4$ ($v_{\rm d}$) & $-9.31 \times 10^{-3}$ & $3.41 \times 10^{-4}$ & -27.30 & -0.0014\\ 
 \hline
\multicolumn{5}{c}{$\sigma = 9.9 \times 10^{-4}$ $M_{\sun}$; $R^2$ = 0.992}\\
  \hline
\end{tabular}
\tablefoot{The $p$-values of the tests are $<2 \times
  10^{-16}$, if not differently specified. The column legend is the same as
in  Table~\ref{table:modLTO}.\\ 
\tablefoottext{*}{$p$-value $= 3.36 \times 10^{-6}$.}\\ }
\end{table}

\section{Conclusions}
\label{sec:conclusions}

In this paper we addressed the problem of a quantitative and systematic
evaluation of the cumulative propagation of physical uncertainties in current
generation of stellar models of low mass stars from the main sequence to the
zero age horizontal branch. At variance with several previous work
\citep[see
  e.g.][]{chaboyer1995,Cassisi1998,Castellani1999,Castellani2000,
  Imbriani2001,PradaMoroni2002,Salaris2002b,Imbriani2004,Weiss2005,
PradaMoroni2007,cefeidi,    
  Tognelli2011}, where a single input physics was changed at a time, we
performed a systematic and simultaneous variation on a fixed grid within their
current range of uncertainty, in a way to obtain a full crossing of the
perturbed input values, of the main physical inputs adopted in stellar
codes (i.e. $^{1}$H(p,$\nu e^+$)$^{2}$H, $^{14}$N$(p,\gamma)^{15}$O, and
triple-$\alpha$ reaction rates, radiative and conductive opacities, neutrino
energy losses, and microscopic diffusion velocities).

Although very expensive from the computational point of view, such an approach
has the important advantage with respect to the previous one to be more
robust against possible interactions among the varied input physics, as any
a priori independence among them is assumed.

Relying on a set of stellar models fully covering all the possible combinations 
of simultaneously perturbed input physics, we were able to compute the error 
stripe associated to our reference stellar track, i.e.  $M = 0.90$ $M_{\sun}$ 
with initial metallicity $Z = 0.006$ and helium abundance $Y = 0.26$, 
  from the pre-main sequence up  
to the RGB tip, in different planes (i.e.  $\log L$ vs $\log T_{\rm eff}$,
$X_{c}$ vs $age$, $\Delta \log L$ vs $model$,  
and $\Delta \log R$ vs $model$). We built also the error stripe associated to
the ZAHB locus  
resulting from the evolution of a progenitor of $M = 0.90$ $M_{\sun}$.
As far as we know, this is the first time that an error stripe is computed and
plotted for stellar tracks (see Appendix \ref{app:nastro} for a detailed
description of the error stripe construction method).

Furthermore, we quantified the extension of the global variability regions
 (i.e. due to the cumulative effect of all physical uncertainties) for some
relevant stellar quantities, without any assumption on the parent
distributions of the varied 
  physical inputs. In particular, we focused on the turn-off luminosity
  $L_{\rm BTO}$,  
the central hydrogen exhaustion time $t_{\rm H}$, the luminosity $L_{\rm tip}$ 
and the helium core mass $M_{\rm c}^{\rm He}$ at the RGB tip, and the ZAHB
luminosity  in the RR Lyrae region $L_{\rm HB}$, were computed, too (see
e.g. Table 
\ref{table:incertezzetracce}). 
We found that the turn-off log luminosity $\log L_{\rm BTO}$ varies of
$\pm 0.021$ dex,   
while the RGB tip $\log L_{\rm tip}$ and ZAHB $\log L_{\rm HB}$ ones
of $\pm 0.03$ dex and $\pm 0.045$ dex, respectively. 
Thus, uncertainties of the order of $\pm 0.075$ mag and $\pm 0.10$ mag should
be taken into  account in distance  
modulo estimates obtained by theoretically calibrated RGB tip and ZAHB
luminosities. 
The predicted central hydrogen exhaustion time $t_{\rm H}$ varies of
$\pm 0.72$ Gyr, whereas 
 the helium core mass at the RGB tip $M_{\rm c}^{\rm He}$ of $\pm 0.0042$
 $M_{\sun}$.   
 
This large computational effort (i.e. 3159 stellar tracks) made possible a
thorough and rigorous statistical analysis of the effects of the variations of
the quoted physical 
inputs on the chosen  relevant stellar evolutionary features, allowing to
disentangle the 
contributions of the different inputs physics.
    
The results of our extended statistical analysis show that the radiative
opacity is, by far, the dominant source of physical uncertainty in almost all
of the examined stellar evolutionary features, with the exception of
 the helium core mass $M_{\rm c}^{\rm He}$ at the RGB
tip which is affected mainly by the uncertainty in the triple-$\alpha$
reaction rate.  
As an example, for the turn-off log luminosity $\log L_{\rm BTO}$, in order to
reduce the impact  
of the radiative opacity variation to the level of the second most important
input, 
i.e. the $^{14}$N(p,$\gamma$)$^{15}$O reaction rate, the uncertainty of the
former should decrease  
from the current 5\% down to 1\%. For the central hydrogen exhaustion time
$t_{\rm H}$,  
 the prevalence of the radiative opacity is even larger, as for reducing its
 effect  
 at the level of the second most important input physics, i. e. the
 microscopic diffusion velocities,  
its uncertainty should be decreased at the 0.56\% level. Notice that the 5\%
uncertainty in radiative opacity  
assumed in our calculations is probably an underestimate, at least in some
regions of the temperature-density  
plane of interest for stellar models. 

Beside this analysis of the cumulative uncertainty in a stellar track of fixed
mass and chemical composition, we extended the previous
 analysis of the cumulative uncertainty due to the main input physics to
 theoretical   
isochrones in the age range 8-14 Gyr. We focused in particular on
the turn off luminosity $L_{\rm BTO}^{\rm iso}$ 
 and mass $M_{\rm BTO}^{\rm iso}$, and 
$\log L_{\rm HB}/\log L_{\rm BTO}^{\rm iso}$,  which is the theoretical 
counterpart of the visual magnitude difference between 
turn-off and horizontal branch regions (i.e. $ \Delta V(TO-HB)$ used as age
indicator in the ``vertical method''). 
For an age of 12 Gyr, we found that the isochrone turn-off log luminosity
$\log L_{\rm BTO}^{\rm iso}$  
varies of $\pm 0.013$ dex, about 2/3 of the value found for the turn-off
luminosity $L_{\rm BTO}$ of the 
reference stellar track. For the same age, the mass at the isochrone turn-off
varies of $\pm 0.015$ $M_{\sun}$ and  
$\log L_{\rm HB}/\log L_{\rm BTO}^{\rm iso}$ of  $\pm 0.05$ dex. The large set of
perturbed isochrones allowed us to perform
the same kind of statistical analysis previously applied to tracks of fixed
mass, with the aim to evaluate the effects  
of the single varied input physics.  

Finally, the availability of isochrones computed by varying simultaneously all
the main input physics made possible 
 to evaluate the physical uncertainty affecting the age inferred from two of
 the most used cosmic clocks, namely the 
 turn-off luminosity and the vertical method. For given $L_{\rm BTO}^{\rm iso}$ and
  $\log L_{\rm HB}/\log L_{\rm BTO}^{\rm iso}$ values, the inferred age varies
 in a of about $\pm$ 0.375 Gyr and $\pm$ 1.25 Gyr, respectively.

An equally systematic approach is provided by Monte Carlo simulations as those 
presented by Chaboyer and collaborators \citep[][]{Chaboyer1996,Chaboyer1998,
  Chaboyer2002,  
 Krauss2003, Bjork2006}. However, a direct comparison with their results is
difficult since  
 the studies address different questions about the global
uncertainty. 
The main problem is due to the fact that we focus on the
uncertainty  
of evolutionary features different from the one studied in those papers.
The only possible feature we can compare is the change in the evolutionary
age due to the variation of radiative opacity. In \citet{Chaboyer2002} a 2\%
change in radiative opacity accounts for a 2.6\% variation (i.e.
6.5\% for a variation of radiative opacity of 5\%)
of the age of a Sub-Giant Branch star. As proxy of this quantity, we found
that a 
5\% increase in radiative opacity account for a change of the time of central
hydrogen exhaustion of 5.5\%, i.e. 1\% lower to the equivalent variation
quoted above. 
The small difference can be in principle due to the different sampling schema
for the radiative opacity adopted in \citet{Chaboyer2002}, and to the different
chemical inputs and 
solar mixture employed in the calculations.

Although the present paper addresses some fundamental topics about the
uncertainty of 
modern stellar models, many other questions remain open and need further
 investigations. 
The first problem concerns the possibility to extend the results of our analysis
-- performed at fixed values of metallicity and initial helium abundance -- to
different values of the chemical inputs. In the lack of the very huge set of
computations required to specifically address this topic, the 
extrapolation of the results presented in this paper to values of $Z$ and $Y$
very 
different from the ones employed in the computations is to be considered with
care. Some cautions should also be adopted for ``borderline'' stellar models
which can develop a different structure -- such as the development of a
convective core -- due to a perturbation of a physical input. In these few
cases the evaluation of the uncertainty can not be inferred but has to be
performed with direct 
computations.  

A second question concerns the possibility to extend the results 
presented here to the computations performed with other stellar evolutionary
codes. This point could be 
addressed only with the replication with other codes, with the very
same inputs and configurations, of the computations
presented here. The availability of such calculations by different groups of the
stellar evolution community would be of invaluable
importance. In fact in this way could be assessed not only the uncertainty due
to the variations of the 
physical inputs, but also to quantify the impact of another source of
uncertainty, i.e. the possible systematic difference among various codes due
to different implementations of physical mechanisms and to the algorithms used
in the computations.

\begin{acknowledgements}
We warmly thank Alexander Potekhin for providing us an estimate of the
uncertainty affecting the electron conduction opacity coefficients. 
We thank Dario Andrea Bini for enlightening suggestions about the error stripe
construction and Steven N. Shore for carefully reading the paper and for
useful comments. 
We are grateful to our anonymous referee for many stimulating suggestions that
helped in clarify and improve the paper.
This work
has been supported by PRIN-INAF 2011 
 ({\em Tracing the formation and evolution of the Galactic Halo with VST}, PI
M. Marconi).  
\end{acknowledgements}

\bibliographystyle{aa}
\bibliography{biblio}

\appendix
\section{Tracks reduction and error stripe construction on the theoretical
  plane}\label{app:nastro} 

The first step of the construction of the  stellar
track uncertainty stripe on the theoretical plane is the reduction of the raw
tracks to a set of  
tracks with the same number $n$ of homologous points. 
The reduction is based upon the identification of some
keystone evolutionary points on the raw track and subsequent deployment of the
same number of points -- obtained by interpolation -- on all the tracks.    

As reference points we adopted the following:
\begin{enumerate}
\item{PMS1: the point for which gravitational luminosity reaches
  0.996 times the surface luminosity.} 
\item{PMS2: the point for which the gravitational luminosity
  decreases of 0.05 from the value at PMS1.} 
\item{ZAMS: the point for which the central hydrogen abundance drops
below 99\% of its initial value.}
\item{MS1: the point for which the central hydrogen abundance drops
below 7\%.} 
\item{MS2: the point for which the central hydrogen abundance drops
below 1\%.} 
\item{MS3: the point for which the central hydrogen abundance drops
below 0.1\%.} 
\item{HC: the point for which the central hydrogen is exhausted.}
\item{RGB1, or RGB start: the point on the track at maximum
  distance from the line connecting the HC and the RGB2 point.}
\item{RGB2, or RGB bump: the point for which the luminosity of RGB
  start decreasing. } 
\item{RGB3, or He-flash: the point for which the He
  burning luminosity reaches 100 times the surface luminosity.}
\end{enumerate}

The ZAHB models have been computed through a synthetic method where the
starting models were  
obtained by accreting envelopes of different mass extensions onto the He-core
left at the tip of the RGB.  
Then, full evolutionary calculations were started again as thermal relaxed
models in the central He  
burning phase; ZAHB point corresponds to the model in which the equilibrium
abundance of CNO burning secondary 
 elements is reached, after about 1 Myr. 

Let $\{ T_i \}$, $i=1,\ldots,N$ be the set of the $N$ reduced tracks and let
$T_i(j)$ be the $j$th point on the $i$th reduced track. Let's
define:
\begin{equation} 
{\cal T}(j) = \{ T_i(j) \} \quad i=1,\ldots,N 
\end{equation}
the set of the $j$th points over the whole set of reduced tracks.
Let $H_{c} (X)$ be the convex hull of a generic set $X$.
We define: 
\begin{equation}
{\cal H}(k) =  H_{c}  \left( \bigcup_{j=k}^{k+1}{{\cal T}(j)} \right)
\end{equation}
the convex hull of the set composed by the $k$th and $(k+1)$th points of the
reduced tracks.   

The full stripe is then given by:
\begin{equation}
{\cal H} = \bigcup_{k=1}^{n-1} {\cal H}(k)
\end{equation}

The required computations were carried out using R 2.14.1 \citep{R}.

\section{Reverse inference on the age of a isochrone given the TO luminosity}
\label{app:intconf}

Let us consider a simple linear model:
\begin{equation}
y = \beta_0 + \beta_1 \; x 
\end{equation}
the 95\% confidence interval ($y_1$, $y_2$) on a future observation $y$ given
$x$, is \citep[see
  e.g.][]{linmodR}:
\begin{equation}
y_{1,2} = \hat{\beta_0} + \hat{\beta_1} \; x \pm t_{n-2}^{1-\alpha/2} \;
\hat{\sigma} \sqrt{1 + \frac{1}{n} + \frac{(x - \bar x)^2}{d_x^2}} 
\end{equation}
where $\hat{\beta_0}$, $\hat{\beta_1}$, $\hat{\sigma}$ are the least-squares
estimates of the model parameters, $\alpha$ is the required confidence level
(in our case, $\alpha$ = 0.05), $n$ is the number of points in the model,
$t_{n-2}^{1-\alpha/2}$ is the $1-\alpha/2$
quantile of the Student $t$ distribution with $n-2$ degrees of freedom,
$\bar x$ is the sample mean value of $x$, $d_x^2$ is the sample deviance of
$x$.

\begin{figure*}
\centering
\includegraphics[width=6.3cm,angle=-90]{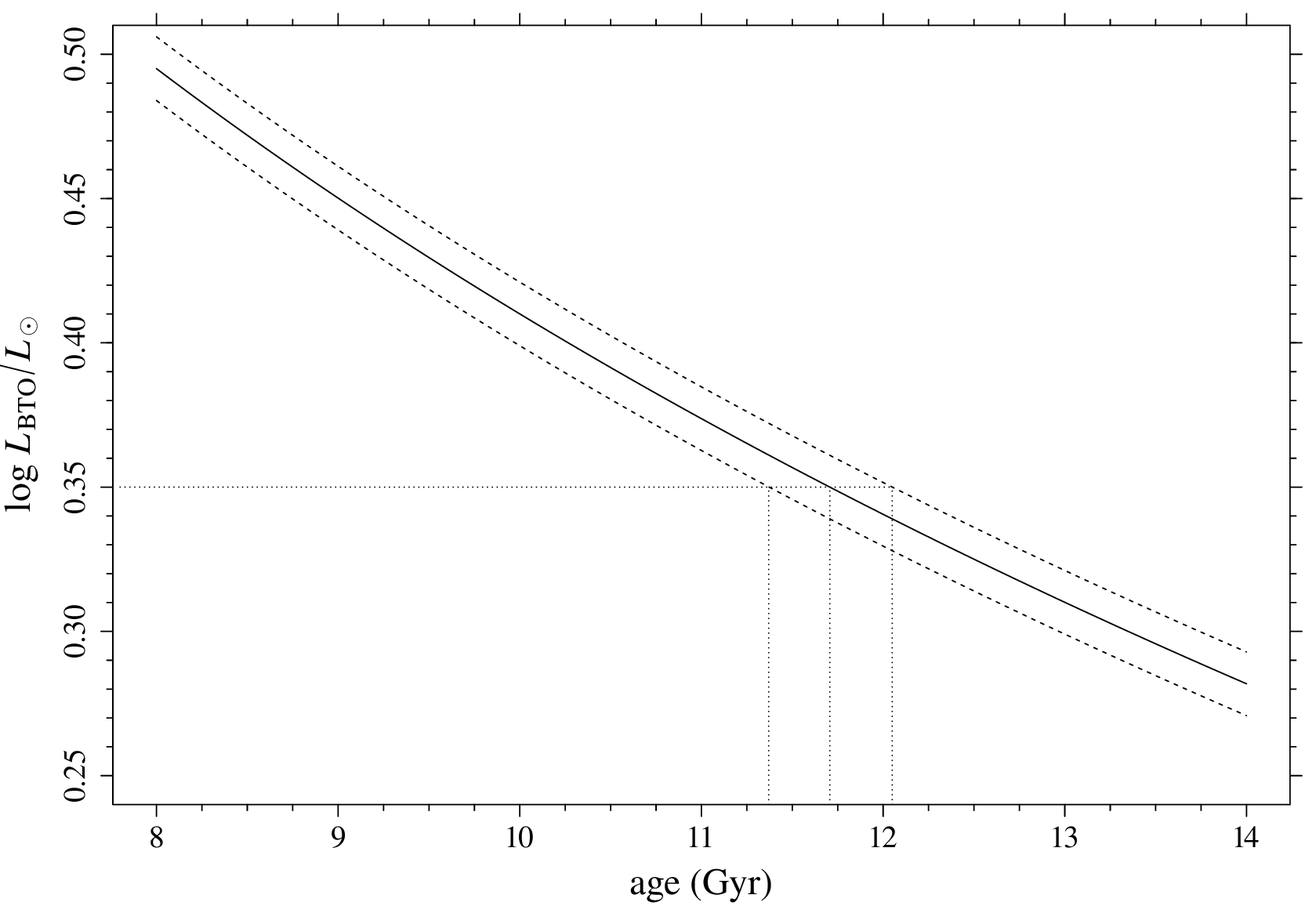}
\includegraphics[width=6.3cm,angle=-90]{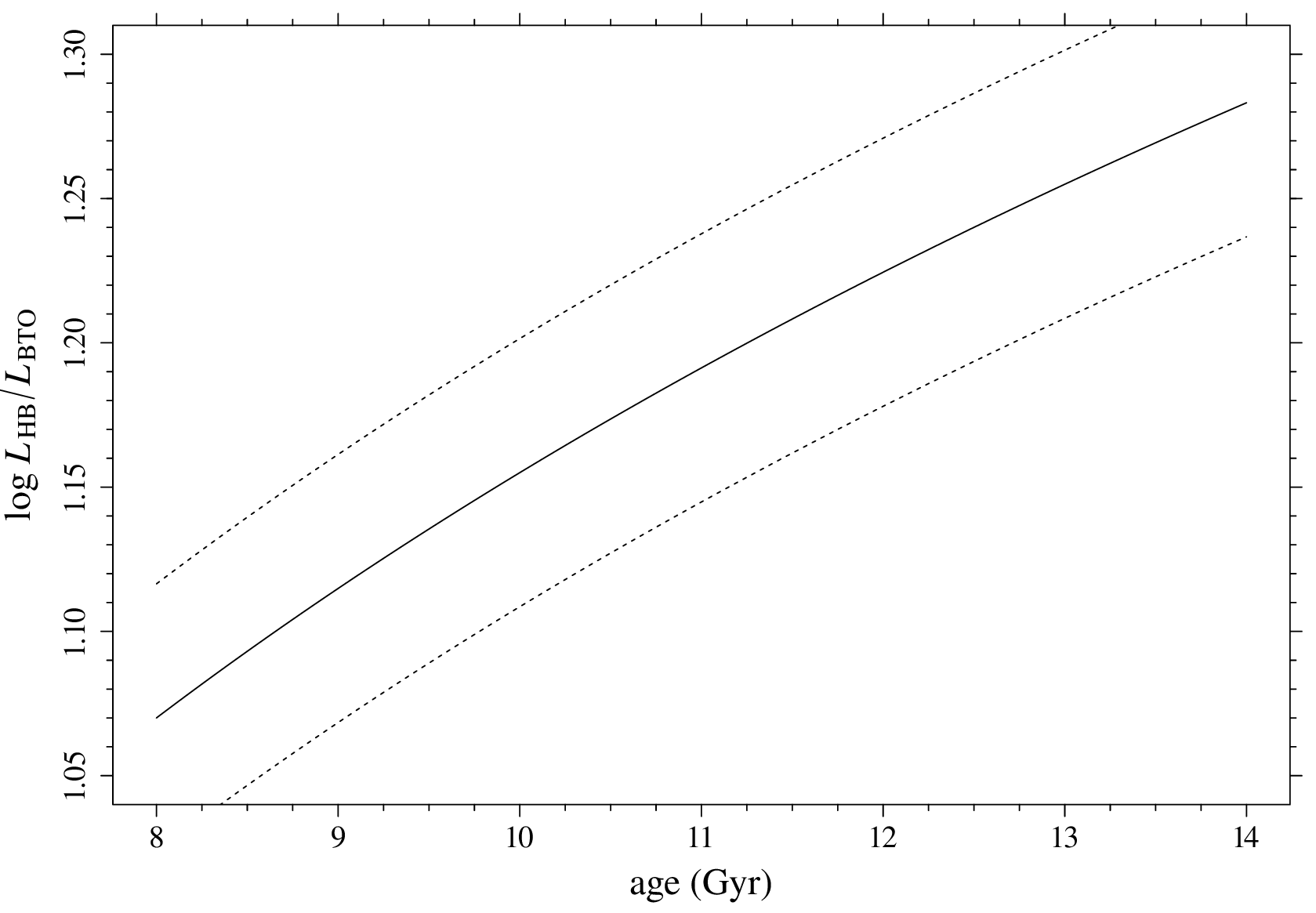}
\caption{Left panel: reverse inference on the isochrone age given the BTO log
  luminosity. The solid line represents the best fit model, the dashed lines
  define the 95\% confidence interval on the prediction of the model. The
  dotted lines show the uncertainty on age given a value of BTO log
  luminosity. Right panel: same for $\log L_{\rm HB}/L_{BTO}^{iso}$.}
\label{fig:revinf}
\end{figure*}

In our case we have: $y = \log L_{\rm BTO}^{\rm iso}$ and $x = \log T$,
with $T$ the age of the isochrone in Gyr.
The boundaries of the 95\% confidence interval are displayed along with the best
fit line in Fig.~\ref{fig:revinf}, where we show the construction of the range
of uncertainty on age, given the value of BTO log luminosity. As an example,
for $\log L_{\rm BTO}^{\rm iso} = 0.35$ dex the estimated ages from the models
lie in the range [11.37 - 12.05] Gyr.

\section{On the mixing-length influence}
\label{app:mixlen}

The results quoted in the present paper are computed for a fixed value of
mixing-length parameter, i.e. $\alpha_{\rm ml}$ = 1.90. The possibility to
extend them to different values of this parameter must be checked by computing
stellar models with different $\alpha_{\rm ml}$.

In presence of an effect due to the mixing-length, we expect a difference
in the regression coefficients for the models summarized in
Tables~\ref{table:modLTO}-\ref{table:modHB383}  
when they are computed from stellar model with different $\alpha_{\rm ml}$. 
The computation of the huge sets of stellar models for different
mixing-length values is not fully needed since in the linear models presented in
Sec.~\ref{sec:statistic_track}  is shown that the
various physical inputs do not interact.
Therefore a subset of stellar models carefully selected can suffice to assess
the presence of a mixing-length effect of distortion on the regression
coefficients.

To perform the analysis we computed, for the two values of the mixing-length
parameter $\alpha_{\rm ml}$ =
1.70, 1.80 ($\alpha_{\rm ml}$ = 1.74 is the solar-calibrated value of the
mixing-length parameter), 81 stellar models each. The models  to be computed
were randomly 
selected 
using a latin hypercube sampling design, which is an extension of
latin square to higher dimensions and has optimal property in reducing
the variance of the estimators obtained from the linear models
\citep{Stein1987}. 
The random selection was performed using the R library lhs \citep{LHS}.
The corresponding 81 models for $\alpha_{\rm ml}$ = 1.90 were extracted from
the 2187 original calculations and these 243 stellar models are used for
establishing the influence of $\alpha_{\rm ml}$.

The analysis is performed by adapting to data linear models of the form:
\begin{equation}
\log L_{\rm BTO}, t_{\rm H}  = \beta_0 + (\sum_{i=1}^4 \beta_i \; p_i) * \alpha_{\rm ml}
\end{equation}
or:
\begin{equation}
\log L_{\rm tip}, \log L_{\rm HB}, M_{\rm c}^{\rm He}  = \beta_0 + 
(\sum_{i=1}^7 \beta_i \; p_i) * \alpha_{\rm ml}
\end{equation}  
where, for convenience, we adopted the operator ``*'' defined as $A * B \equiv
A + B + A \cdot B$.
The effect of distortion of the regression coefficients due to the presence of
the mixing-length parameter is represented by the interaction between each
multiplier $p_i$ and $\alpha_{\rm ml}$. 
The statistical significance of these coefficients is of limited importance,
since it can be arbitrarily increased 
by selecting a larger subsample thus reducing the standard error estimate of
the coefficients. A more useful indicator is the physical impact of the
introduction of the interaction in the linear models. As explained in the text
the impact of a perturbation $\Delta p_i$ on a parameter $p_i$ which enters a
linear model can be
evaluated as $\Delta p_i \cdot \beta_i$, where $\beta_i$ is the estimate of  the
regression coefficient. In presence of the interaction the impact of the same
perturbation $\Delta p_i$ combined with a variation of $\Delta \alpha_{\rm
  ml}$ can be estimated as $\Delta p_i \cdot \Delta \alpha_{\rm
  ml} \cdot \beta_j$ where $\beta_j$ is the regression coefficient of the
interaction term under analysis. In table \ref{table:mixlen} we report the
impact of the interaction terms in the various models for $\Delta \alpha_{\rm
  ml} = 0.1$. In all cases the physical impact of the interaction is
negligible, so we can conclude that the regression presented in
Sec.~\ref{sec:statistic_track} are robust for an acceptable change in
$\alpha_{\rm ml}$.

\begin{table}[ht]
\centering
\caption{Additional impact of the variation of the mixing-length value by 0.1
  on the physical impacts presented in Sec.~\ref{sec:statistic_track}.}
\label{table:mixlen}
\centering
\begin{tabular}{lrrrrr}
  \hline\hline
 source & $\log L_{\rm BTO}$ & $t_{\rm H}$ & $\log L_{\rm tip}$ & $M_{\rm
  c}^{\rm He}$ & $\log L_{\rm HB}$ \\
   & (dex)  & (Gyr) & (dex) & ($M_{\sun}$) & (dex)\\ 
\hline 
 $\alpha_{\rm ml} \cdot {\rm pp}$ &  0.00002 &  0.0008 &  -0.00006 &  $1.3
\times 10^{-6}$ &   0.00004\\
 $\alpha_{\rm ml} \cdot ^{14}$N    &  0.00002 &  0.0000 &  0.00002  &  $2.1
\times 10^{-7}$ &  -0.00001\\
 $\alpha_{\rm ml} \cdot k_{\rm r}$  &  0.00026 &  0.0033 &  0.00002  &
-$6.0 \times 10^{-6}$ &  -0.00006\\
 $\alpha_{\rm ml} \cdot v_{\rm d}$  &  0.00000 &  0.0001 &  0.00000  &  $2.7
\times 10^{-7}$ &   0.00001\\
 $\alpha_{\rm ml} \cdot 3\alpha$  &    --     &  --      & -0.00016  & -$4.0
\times 10^{-6}$ &  -0.00009\\
 $\alpha_{\rm ml} \cdot \nu$      &     --    &   --     & 0.00014   & $1.7
\times 10^{-6}$ &   0.00002\\
 $\alpha_{\rm ml} \cdot k_{\rm c}$  &    --     &   --     & -0.00006  & -$5.4
\times 10^{-7}$ &  -0.00008\\
\hline
\end{tabular}
\end{table}

\section{On the EOS influence}
\label{app:eos}

The assessment of the EOS influence on the stellar models can not be done
in the same way of the other physical inputs, given the available information.
 As discussed in the text, the thermodynamic quantities required for computing
stellar models and provided by EOS are related  together
in a not trivial way, hampering a simple parametrization of the uncertainty
associated with them.

Nevertheless, a rough estimate of the impact due to the present
uncertainty on EOS can be computed in an alternative way. 
For this purpose, in this Appendix we provide
the output of the stellar models computed with two different and widely
adopted choices for
equation-of-state: OPAL EOS and FreeEOS \citep{irwin04}.   

\begin{table}[ht]
\centering
\caption{Impact of equation of state change from OPAL to FreeEOS for the
 examined evolutionary features. }
\label{table:eos}
\centering
\begin{tabular}{lrrr}
  \hline\hline
Evolutionary feature & OPAL & FreeEOS & Difference\\
\hline 
$\log L_{\rm BTO}$  (dex)        & 0.3548 & 0.3520 & 0.0028  \\
$t_{\rm H}$ (Gyr) & 10.557 & 10.461 & 0.096  \\
$\log L_{\rm tip}$ (dex)    & 3.4055 & 3.4075 & -0.0020 \\
$M_{\rm c}^{\rm He}$ ($M_{\sun}$)         & 0.4835 & 0.4850 & -0.0015 \\ 
$\log L_{\rm HB}$ (dex)       & 1.5634 & 1.5704 & -0.0070 \\
\hline
\end{tabular}
\tablefoot{First column: evolutionary stage; second
 column: values obtained with OPAL EOS; third column: values obtained with
 FreeEOS; last column: differences among the two values.}
\end{table}

Table~\ref{table:eos}
reports -- for each evolutionary feature -- the values obtained using the
two different EOS and, in the last column, the impact of the
EOS change. By comparing them with the values in the last column of
Table~\ref{table:incertezzetracce}, 
one sees that the EOS change accounts for about 1/7 - 1/8 of the
total variation  due to all the other physical inputs.
The only exception is the He core mass, for which the EOS change accounts
for 1/3 of the total variation.

Unlike the other analysis performed in the paper, in this case it is not
possible to identify which of the varied thermodynal quantities mainly affects
the total result. One also should be aware that a given thermodynamical
quantity can have different influences in different evolutionary phases.

\end{document}